%% file: main.tex
\documentclass[lettersize,journal]{IEEEtran}
\IEEEoverridecommandlockouts
\usepackage[noadjust]{cite}
\usepackage{amsmath,amssymb,amsfonts}
\usepackage{algorithmic}
\usepackage{graphicx}
\usepackage{textcomp,dsfont}
\usepackage{xcolor,comment}
\usepackage{nicefrac,xfrac}
\usepackage{subcaption}
\usepackage{flushend}

\newtheorem{theorem}{Theorem}

\newtheorem{definition}{Definition}

\def\BibTeX{{\rm B\kern-.05em{\sc i\kern-.025em b}\kern-.08em
    T\kern-.1667em\lower.7ex\hbox{E}\kern-.125emX}}

\begin{document}

\title{Instantaneous Polarimetry with Zak-OTFS}
% \thanks{
% This work is supported by the National Science Foundation under grants 2342690 and 2148212, in part by funds from federal agency and industry partners as specified in the Resilient \& Intelligent NextG Systems (RINGS) program, and in part by the Air Force Office of Scientific Research under grants FA 8750-20-2-0504 and FA 9550-23-1-0249. \\
% $*$ denotes equal contribution.}
% }

\author{Nishant Mehrotra$^*$, Sandesh Rao Mattu$^*$, Robert Calderbank~\IEEEmembership{Life Fellow,~IEEE}
        % <-this % stops a space
\thanks{This work is supported by the National Science Foundation under grants 2342690 and 2148212, in part by funds from federal agency and industry partners as specified in the Resilient \& Intelligent NextG Systems (RINGS) program, and in part by the Air Force Office of Scientific Research under grants FA 8750-20-2-0504 and FA 9550-23-1-0249. \\ The authors are with the Department of Electrical and Computer Engineering, Duke University, Durham, NC, 27708, USA (email: \{nishant.mehrotra,\allowbreak sandesh.mattu,\allowbreak robert.calderbank\}\allowbreak@duke.edu). \\ $*$ denotes equal contribution.}% <-this % stops a space
%\thanks{Manuscript received April 19, 2021; revised August 16, 2021.}
}

% \author{\IEEEauthorblockN{Nishant Mehrotra$^*$}
% \IEEEauthorblockA{\textit{Electrical and Computer Engineering} \\
% \textit{Duke University}\\
% Durham, USA \\
% nishant.mehrotra@duke.edu\vspace{0mm}}
% \and
% \IEEEauthorblockN{Sandesh Rao Mattu$^*$}
% \IEEEauthorblockA{\textit{Electrical and Computer Engineering} \\
% \textit{Duke University}\\
% Durham, USA \\
% sandesh.mattu@duke.edu\vspace{0mm}}
% \and
% \IEEEauthorblockN{Robert Calderbank}
% \IEEEauthorblockA{\textit{Electrical and Computer Engineering} \\
% \textit{Duke University}\\
% Durham, USA \\
% robert.calderbank@duke.edu\vspace{0mm}}
% }

\maketitle
\begin{abstract}
Polarimetry, which is the ability to measure the scattering response of the environment across orthogonal polarizations, is fundamental to enhancing wireless communication and radar system performance. In this paper, we utilize the Zak-OTFS modulation to enable \emph{instantaneous} polarimetry within a single transmission frame. We transmit a Zak-OTFS carrier waveform and a spread carrier waveform mutually unbiased to it simultaneously over orthogonal polarizations. The mutual unbiasedness of the two waveforms enables the receiver to estimate the full polarimetric response of the scattering environment from a single received frame. Unlike existing methods for instantaneous polarimetry with computational complexity quadratic in the time-bandwidth product, the proposed method enables instantaneous polarimetry at \textcolor{black}{near-linear} complexity in the time-bandwidth product. Via numerical simulations, we show ideal polarimetric target detection and parameter estimation results with the proposed method, with improvements in computational complexity \textcolor{black}{and greater clutter resilience} over comparable baselines.
\end{abstract}


\input{intro}

\input{prelim}

\input{pol_radar}

\input{results}

\input{conclusion}

\bibliographystyle{IEEEtran}
\bibliography{references}
\end{document}

%% file: intro.tex
\section{Introduction}
\label{sec:intro}

\IEEEPARstart{P}{olarimetry} is an important tool for enhancing the performance of both wireless communication and radar systems. In wireless communication, polarimetry provides a diversity gain~\cite{Vaughan1990_pol_div,Paulraj2002_pol_div,Valuenzela2002_pol_div,Mark2006_pol_div}, thereby improving the reliability of communication, as well as a spatial multiplexing gain~\cite{Andrews2001_pol_dof,Marzetta2002_pol_dof,Hughes2008_pol_dof,Poon2011_pol_dof}, which increases the capacity of the wireless link. Similarly, polarimetry increases the waveform degrees-of-freedom in radar systems~\cite{Boerner1990_polsurvey,Nehorai2009_polsurvey,Calderbank2009_wvfsurvey,Antar2002_polcomparison,Giuli1990_polsimult1,howard2007simple,Pezeshki2008,Calderbank2006_pol_phasecoded,Hochwald1995_polmodel,Cloude2005_poleig}, providing more information about the target and enabling improved detection of targets with small radar cross section (RCS), such as drones.

\begin{figure*}[!t]
    \centering
    \begin{subfigure}{0.43\linewidth}
        \centering
        \includegraphics[width=\linewidth]{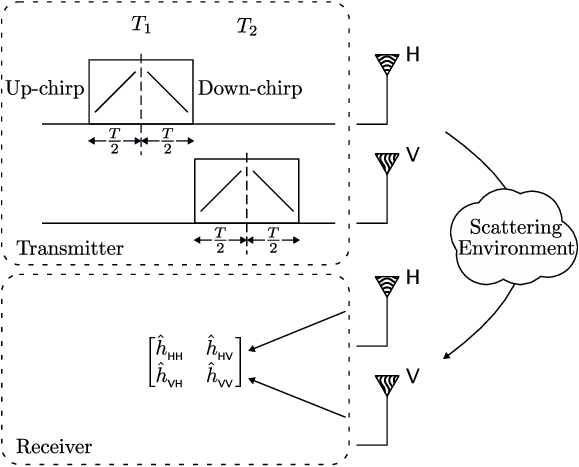}
        \caption{Sequential Polarimetry with FMCW}
        \label{fig:pol_fmcw}
    \end{subfigure}
    \hspace{15mm}
    \begin{subfigure}{0.39\linewidth}
        \centering
        \includegraphics[width=\linewidth]{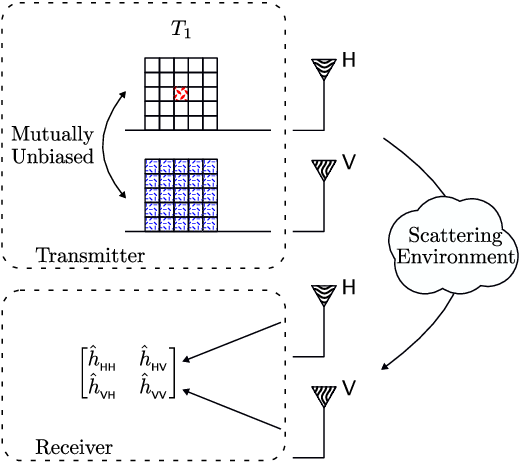}
        \caption{Instantaneous Polarimetry with Zak-OTFS (Ours)}
        \label{fig:pol_zak}
    \end{subfigure}
    \caption{Comparison of different approaches for polarimetry. (a) Sequential polarimetry with FMCW transmits polarized FMCW waveforms over two frames, with each frame subdivided into two halves with an up-chirp and a down-chirp respectively. The associated Doppler resolution is $\nicefrac{2}{T}$ and the computational complexity is $\mathcal{O}(B^2T^2)$. (b) Instantaneous polarimetry with Zak-OTFS transmits a Zak-OTFS pulsone and a mutually unbiased spread waveform obtained via a unitary transformation of the pulsone in a single frame. Compared to the sequential approach in (a), the proposed approach has $2 \times$ smaller latency, $2 \times$ improved Doppler resolution of $\nicefrac{1}{T}$, and a computational complexity of only $\mathcal{O}(BT \log T)$.}
    \label{fig:pol_block_dia}
\end{figure*}

Polarimetry is enabled in radar and communication systems by transmitting and receiving on two orthogonal polarizations, e.g., on vertical and horizontal polarizations. The receiver estimates the $2 \times 2$ \emph{polarimetric scattering response} of the wireless/radar channel across all four combinations of transmit and receive polarizations. A standard approach is to transmit polarized waveforms \emph{sequentially} across two frames~\cite{Calderbank2009_wvfsurvey,Antar2002_polcomparison,Giuli1990_polsimult1,howard2007simple,Pezeshki2008}; see Fig.~\ref{fig:pol_block_dia}(\subref{fig:pol_fmcw}) for an example with frequency modulated continuous wave (FMCW) transmissions. From its measurements in each frame, the receiver estimates $2 \times 1$ slices of the full $2 \times 2$ polarimetric scattering response. Such an approach does not provide instantaneous estimates of the scattering response within a single frame. Changes in the scattering environment between the two frames (due to mobility) may partially decorrelate the obtained estimates~\cite{Antar2002_polcomparison,Giuli1990_polsimult1,howard2007simple,Pezeshki2008,Calderbank2006_pol_phasecoded}. Sequential polarimetry also prevents frame-by-frame processing \& increases the system latency, which is a critical factor for radar and communication performance in highly dynamic environments. \textcolor{black}{When utilizing continuous waveforms, such as FMCW and pulsed waveforms~\cite{Jankiraman2018_fmcw,Uysal2020_phasecoded_fmcw,Skolnik1980,Levanon2004}, the computational complexity of sequential polarimetry is \emph{quadratic} in the time-bandwidth product~\cite{Calderbank2015_ltv,zakotfs_ltv,EURASIP2025}}.

% \begin{figure*} 
%     \centering
%     % \hspace{10mm}
%     \subfloat[{Sequential Polarimetry with FMCW}]{\includegraphics[width=0.43\linewidth]{polarimetry_fmcw.eps}\label{fig:pol_fmcw}}
%     \hspace{15mm}
%     \subfloat[{Instantaneous Polarimetry with Zak-OTFS (Ours)}]{\includegraphics[width=0.39\linewidth]{polarimetry_zak.eps}\label{fig:pol_zak}}
%     \hfill
%     \caption{Comparison of different approaches for polarimetry. (a) Sequential polarimetry with FMCW transmits polarized FMCW waveforms over two frames, with each frame subdivided into two halves with an up-chirp and a down-chirp respectively. The associated Doppler resolution is $\nicefrac{2}{T}$ and the computational complexity is $\mathcal{O}(B^2T^2)$. (b) Instantaneous polarimetry with Zak-OTFS transmits a Zak-OTFS pulsone and a mutually unbiased spread waveform obtained via a unitary transformation of the pulsone in a single frame. Compared to the sequential approach in (a), the proposed approach has $2 \times$ smaller latency, $2 \times$ improved Doppler resolution of $\nicefrac{1}{T}$, and a computational complexity of only $\mathcal{O}(BT \log T)$.}
%     % \vspace{-4mm}
%     \label{fig:pol_block_dia}
% \end{figure*}

To unlock the full benefits of polarimetry, it is crucial to estimate the $2 \times 2$ polarimetric scattering response \emph{instantaneously} within a single transmission frame. Previous work~\cite{Giuli1990_polsimult1,howard2007simple,Pezeshki2008} has proposed transmitting \emph{mutually unbiased} waveforms\footnote{\textcolor{black}{The term ``mutually unbiased'' is from quantum information theory~\cite{Schwinger1960unitary}. Formally, two $d$-length waveforms are mutually unbiased if their inner product has magnitude $\nicefrac{1}{\sqrt{d}}$. Measurements from one waveform are ``statistically independent'' to those from the other waveform with uniform probability $\nicefrac{1}{d}$.}}, i.e., waveforms with small inner products, simultaneously across orthogonal polarizations. Mutual unbiasedness ensures that the contribution of the other waveform looks like noise to the receiver when it projects its measurements onto the basis of one of the transmit waveforms. Projecting its measurements onto the basis of each of the two transmit waveforms provides the receiver with an estimate of different $2 \times 1$ slices of the polarimetric scattering response; thus enabling full $2 \times 2$ polarimetric scattering response estimation from a single received frame. Mutually unbiased waveforms have been designed in prior work~\cite{Giuli1990_polsimult1,howard2007simple,Pezeshki2008} via \emph{phase-coding}, i.e., by modulating a common carrier waveform, e.g., a rectangular waveform, with mutually unbiased sequences, e.g., Zadoff-Chu sequences with distinct roots~\cite{Giuli1990_polsimult1} or complementary Golay pairs~\cite{howard2007simple,Pezeshki2008}. While this approach offers excellent polarimetric target detection and parameter estimation (detailed later in Section~\ref{sec:results}), the computational complexity of polarimetry via phase-coding remains \emph{quadratic} in the time-bandwidth product~\cite{EURASIP2025,Fei2024_phasecoded_compl}. Moreover, we have shown in~\cite{EURASIP2025} that separate selection of sequences and carrier waveforms, as done in phase-coding, may be sub-optimal from a radar waveform design perspective. %\emph{phase-coded waveforms} across the two polarizations in a single frame, with the special property that the between (delay and Doppler-shifted copies of) the two waveforms is a small constant

\begin{table}
    \centering
    \caption{Comparison of different approaches for polarimetry; $B$ denotes signaling bandwidth and $T$ denotes frame interval.} %$P_{\mathsf{d}}$ (resp. $P_{\mathsf{fa}}$) denotes probability of detection (resp. false alarm), $B$ is the signal bandwidth and $T$ is the frame interval.} Zak-OTFS jointly optimizes the discrete sequence \& the carrier waveform at lower complexity than phase-coded (separate sequence vs carrier optimization) and FMCW (only carrier optimization).
    {
    \setlength{\tabcolsep}{2.25pt}
    \renewcommand{\arraystretch}{1.25}
    \begin{tabular}{|c|c|c|c|c|}
         \hline
         Approach & Frame(s) & Target(s) & Doppler Res. & Complexity \\ %Detection Perf. Optimization \\
         \hline
         \textbf{Zak-OTFS (Ours)} & \textbf{1} & $\mathbf{> 1}$ & $\mathbf{\nicefrac{1}{T}}$ & $\mathbf{\mathcal{O}(BT\log T)}$ \\ %\textbf{High $P_{\mathsf{d}}$, low $P_{\mathsf{fa}}$} Seq. $+$ Carrier \\
         Phase-coded~\cite{Giuli1990_polsimult1,howard2007simple,Pezeshki2008} & $1$ & $> 1$ & $\nicefrac{1}{T}$ & $\mathcal{O}(B^2T^2)$ \\ %High $P_{\mathsf{d}}$, low $P_{\mathsf{fa}}$ Seq. $\neq$ Carrier \\
         %\hline
         FMCW~\cite{Antar2002_polcomparison,Jankiraman2018_fmcw} & $2$ & $1$ & $\nicefrac{2}{T}$ & $\mathcal{O}(B^2T^2)$ \\ %Low $P_{\mathsf{d}}$, high $P_{\mathsf{fa}}$ Carrier only \\
         %\hline
         \hline
    \end{tabular}
    }
    \vspace*{-0.1in}
    \label{tab:prior_work}
\end{table}

In this paper, we take an alternate approach to constructing mutually unbiased waveforms for instantaneous polarimetry by utilizing the \emph{Zak-OTFS} (orthogonal time frequency space) modulation~\cite{otfs_book,bitspaper1,bitspaper2} in place of phase-coding. The carrier waveform in Zak-OTFS, termed \emph{pulsone}, is a pulse localized in the delay-Doppler domain. We utilize a generalized discrete affine Fourier transform~\cite{EURASIP2025,Mehrotra2025_WCLSpread} to transform the pulsone into a spread waveform that is mutually unbiased to the pulsone. Our approach has two advantages over phase-coding. First, scattering response estimation with pulsones and their unitary transformations is possible with \emph{\textcolor{black}{near-linear}} computational complexity in the time-bandwidth product~\cite{EURASIP2025}. Second, unlike phase-coding, Zak-OTFS enables \emph{joint optimization} of sequences and carrier waveforms, which has been shown to be optimal from a radar waveform design perspective in~\cite{EURASIP2025}. We illustrate our proposed approach in Fig.~\ref{fig:pol_block_dia}(\subref{fig:pol_zak}) and describe it in more detail in Section~\ref{sec:pol_zak}. %which makes it well-suited for both communication and radar sensing in environments with large delay and Doppler spreads In Zak-OTFS, information is transmitted in the \emph{delay-Doppler} (DD) domain by modulating DD pulses, which are termed \emph{pulsones}. and propose an alternate architecture for instantaneous polarimetry based on the \emph{Zak-OTFS} (orthogonal time frequency space) modulation~\cite{otfs_book,bitspaper1,bitspaper2}. The mutually unbiased pulsone and spread waveforms are transmitted simultaneously over orthogonal polarizations, similar to phase-coding described previously. 

Table~\ref{tab:prior_work} places our contributions in the context of prior work, which is described in more detail in Section~\ref{sec:prelim}. In addition to the advantages in computational complexity and latency, our proposed approach also improves upon the number of simultaneously detectable targets and the Doppler resolution over the sequential approach based on FMCW. Moreover, in Section~\ref{sec:results} we demonstrate that the proposed approach achieves ideal polarimetric target detection and parameter estimation\footnote{\textcolor{black}{The results in this paper serve as a proof-of-concept numerical demonstration of the concept; hardware demonstration will be pursued in future work. Potential challenges include managing the peak-to-average power ratio (PAPR) and over-the-air synchronization. With regards to the former, we note that a natural advantage of the proposed framework is its low PAPR of only $5.6$ dB; see~\cite[Section 4.2]{EURASIP2025} and~\cite{Mehrotra2025_WCLSpread} for details. This reduction in PAPR has also been experimentally demonstrated in~\cite{AFRL2025} in the sub-terahertz band.}} \textcolor{black}{and greater clutter resilience} compared to the phase-coded approach at smaller computational complexity. %\footnote{The results in this paper are limited to polarimetric target detection and parameter estimation in the presence of noise. Extensions to clutter~\cite{Skolnik1980,Spafford1968_clutter,Shnidman1999_clutter,Shnidman2005_clutter,Calderbank2007_seaclutter,Nehorai2009_polsurvey} are possible, but are not pursued in this paper.}

\textit{Notation:} $x$ denotes a complex scalar, $\mathbf{x}$ denotes a vector with $n$th entry $\mathbf{x}[n]$, and $\mathbf{X}$ denotes a matrix with $(n,m)$th entry $\mathbf{X}[n,m]$. $(\cdot)^{\ast}$ denotes complex conjugate, $(\cdot)^{\top}$ denotes transpose, $(\cdot)^{\mathsf{H}}$ denotes complex conjugate transpose and $\langle \mathbf{x}, \mathbf{y} \rangle = \sum_{n} \mathbf{x}[n] \mathbf{y}^{\ast}[n]$ denotes the inner product. Calligraphic font $\mathcal{X}$ denotes operators or sets, with usage clear from context. $\emptyset$ denotes the empty set. $\mathbb{Z}$ denotes the set of integers and $\mathbb{Z}_{N}$ the set of integers modulo $N$. $(a,b)$ denotes the greatest common divisor of two integers $a,b$. $(\cdot)_{{}_{N}}$ denotes the value modulo $N$ and $(\cdot)^{-1}_{{}_{N}}$ denotes the inverse modulo $N$. $\delta(\cdot)$ denotes the delta function, $\delta[\cdot]$ denotes the Kronecker delta function and $\mathbf{I}_{N}$ denotes the $N \times N$ identity matrix. %and $\mathds{1}{\{\cdot\}}$ denotes the indicator function. %, and $\mathbf{e}_{n}$ denotes the standard basis vector with value $1$ at location $n$ and zero elsewhere. $\lfloor \cdot \rfloor$ and $\lceil \cdot \rceil$ denote the floor and ceiling functions

%% file: prelim.tex
\section{Polarimetry: Preliminaries}
\label{sec:prelim}

As described in the Introduction, polarimetry is enabled in radar and communication systems by transmitting and receiving on orthogonal polarizations, e.g., on vertical and horizontal polarizations, using dual-polarized antennas. Let $\mathsf{V}$ and $\mathsf{H}$ respectively denote vertical and horizontal polarization. We now describe how to model the polarimetric scattering response of a $P$-path wireless/radar channel. In uni-polarized systems, the channel gain of each path $p \in \{1,\cdots,P\}$ is modeled by a complex scalar $h^{(p)}$. With dual-polarized transmit and receive antennas, the channel gain is modeled instead by a $2 \times 2$ \emph{polarimetric scattering response}~\cite{Nehorai2009_polsurvey,Calderbank2009_wvfsurvey,Antar2002_polcomparison,Giuli1990_polsimult1,howard2007simple,Pezeshki2008}:
\begin{align}
    \label{eq:pol_prelim1}
    \mathbf{H}^{(p)} &= \begin{bmatrix}
        h^{(p)}_{\mathsf{HH}} & h^{(p)}_{\mathsf{HV}} \\ h^{(p)}_{\mathsf{VH}} & h^{(p)}_{\mathsf{VV}}
    \end{bmatrix} = \mathbf{C}_{\mathsf{RX}} \mathbf{\Sigma}^{(p)} \mathbf{C}_{\mathsf{TX}},
\end{align}
where $\mathbf{C}_{\mathsf{TX}}$ (resp. $\mathbf{C}_{\mathsf{RX}}$) is a $2 \times 2$ matrix characterizing the polarization coupling at the transmitter (resp. receiver), and $\mathbf{\Sigma}^{(p)}$ is a $2 \times 2$ matrix of polarimetric scattering coefficients of the $p$th path/target\footnote{$\mathbf{\Sigma}^{(p)} = \mathbf{I}_{2}$ for a line-of-sight path with no reflection.}. Broadly stated, the goal in polarimetry is to estimate all four components of the polarimetric scattering response in~\eqref{eq:pol_prelim1} from the measurements at the receiver. We now describe two approaches for polarimetry from previous work, before outlining our proposed approach in Section~\ref{sec:pol_zak}.

\subsection{Sequential Polarimetry via FMCW}
\label{subsec:prelim_pol_fmcw}

As described in the Introduction, a standard approach for enabling polarimetry is to transmit polarized waveforms sequentially across two frames. Fig.~\ref{fig:pol_block_dia}(\subref{fig:pol_fmcw}) illustrates sequential polarimetry using FMCW waveforms~\cite{Antar2002_polcomparison,Jankiraman2018_fmcw,Uysal2020_phasecoded_fmcw,Skolnik1980,Levanon2004} (chirps), although the underlying approach is applicable to any waveform. In frame interval $T_1$ (resp. $T_2$), the same waveform is transmitted in horizontal (resp. vertical) polarization. From its measurements in frame interval $T_1$ (resp. $T_2$), the receiver obtains maximum likelihood estimates for $h^{(p)}_{\mathsf{HH}}$ and $h^{(p)}_{\mathsf{VH}}$ (resp. $h^{(p)}_{\mathsf{VH}}$ and $h^{(p)}_{\mathsf{VV}}$) by cross-correlating its dual-polarized received signals and the transmitted waveform~\cite{Antar2002_polcomparison,Jankiraman2018_fmcw,Uysal2020_phasecoded_fmcw,Skolnik1980,Levanon2004}.

% Recall that the receiver's goal is to estimate the $2 \times 2$ matrix $\mathbf{H}^{(p)}$ from~\eqref{eq:pol_prelim1}. For ease of development, assume the scattering environment does not change between the two frames and a single path/target, $P = 1$. From its measurements in each frame, the receiver estimates $2 \times 1$ columns of the $2 \times 2$ matrix $\mathbf{H}^{(p)}$. For example, in frame interval $T_1$ (resp. $T_2$), the receiver can obtain maximum likelihood estimates for $h^{(p)}_{\mathsf{HH}}$ and $h^{(p)}_{\mathsf{VH}}$ (resp. $h^{(p)}_{\mathsf{VH}}$ and $h^{(p)}_{\mathsf{VV}}$) by \emph{cross-correlating} its dual-polarized received signals and the transmitted waveform~\cite{Antar2002_polcomparison,Jankiraman2018_fmcw,Uysal2020_phasecoded_fmcw,Skolnik1980,Levanon2004}.

In the specific case of sequential polarimetry via FMCW, it is known from~\cite{Calderbank2015_ltv,zakotfs_ltv,EURASIP2025} that cross-correlation-based channel estimation has computational complexity $\mathcal{O}(B^2T^2)$, quadratic in the time-bandwidth product $BT$. Localizing a single path/target with FMCW requires \emph{subdividing} each frame interval $T_1$ or $T_2$ into two halves and transmitting a chirp with positive slope (``up-chirp'') and a chirp with negative slope (``down-chirp'') in each half~\cite{Calderbank2015_ltv,zakotfs_ltv}, yielding an effective Doppler resolution of $\nicefrac{2}{T}$. Localizing multiple paths/targets with this approach\footnote{It is possible to localize multiple paths/targets by subdividing each frame into four quarters with further degraded Doppler resolution of $\nicefrac{4}{T}$~\cite{Calderbank2015_ltv,zakotfs_ltv}.} results in multiple false (``ghost'') targets~\cite{Calderbank2015_ltv,zakotfs_ltv}, degrading the detection performance. Moreover, the high sidelobes of FMCW makes detecting weak targets in the presence of stronger ones challenging~\cite{zakotfs_ltv,EURASIP2025}.

There are two primary drawbacks of the sequential approach independent of the drawbacks due to the choice of waveform. First, it requires the scattering environment to remain constant across two frames. Second, it prevents frame-by-frame processing. We now describe an approach for instantaneous polarimetry that overcomes these two drawbacks.

\subsection{Instantaneous Polarimetry via Phase-Coding}
\label{subsec:prelim_pol_phasecoding}

Instantaneous polarimetry overcomes the two drawbacks of sequential polarimetry by transmitting \emph{mutually unbiased} waveforms simultaneously across both polarizations. Fig.~\ref{fig:pol_block_dia}(\subref{fig:pol_zak}) illustrates the main idea. In a single frame interval, unit-norm waveforms $\mathbf{x}_{\mathsf{V}}(t)$ and $\mathbf{x}_{\mathsf{H}}(t)$ satisfying the property $\big|\int \mathbf{x}_{H}(t) \mathbf{x}_{V}^{*}(t-\tau) e^{-j2\pi\nu(t-\tau)} dt\big| \ll 1$ are transmitted in vertical and horizontal polarizations. The receiver estimates all four components of the $2 \times 2$ matrix $\mathbf{H}^{(p)}$ from~\eqref{eq:pol_prelim1} by cross-correlating its dual-polarized received signals with the corresponding transmitted waveform. For example, the receiver estimates $h^{(p)}_{\mathsf{VH}}$ by cross-correlating its received signal in vertical polarization with $\mathbf{x}_{\mathsf{H}}(t)$.

Mutually unbiased waveforms have been designed in previous work~\cite{Giuli1990_polsimult1,howard2007simple,Pezeshki2008} by phase coding a common carrier waveform with mutually unbiased discrete sequences, e.g., Zadoff-Chu sequences with distinct roots. The primary drawback of phase-coding is that the computational complexity of cross-correlation remains quadratic in the time-bandwidth product, $\mathcal{O}(B^2T^2)$~\cite{EURASIP2025,Fei2024_phasecoded_compl}. In the next Section, we describe how to design mutually unbiased waveforms using the Zak-OTFS modulation to enable instantaneous polarimetry with only $\mathcal{O}(BT \log T)$ computational complexity.

%% file: pol_radar.tex
\section{Instantaneous Polarimetry via Zak-OTFS}
\label{sec:pol_zak}

We provide a brief overview of Zak-OTFS in the standard uni-polarized setting in Section~\ref{subsec:prelim_zak}, referring the interested reader to~\cite{otfs_book,bitspaper1,bitspaper2} for a more detailed description of Zak-OTFS. We then extend the system model to polarimetry in Section~\ref{subsec:pol_zak} and detail our proposed approach in Section~\ref{subsec:prop_appr}.

% The architecture of our proposed Zak-OTFS-based polarimetric radar is shown in Fig.~\ref{fig:pol_block_dia}(\subref{fig:pol_zak}). In our proposed approach, we transmit \emph{mutually unbiased} waveforms across both polarizations within a \emph{single} Zak-OTFS frame interval. Each polarized waveform manifests as noise to the other polarization, and hence enables full polarimetric radar processing at the receiver within a single frame with only $\mathcal{O}(BT \log T)$ computational complexity. Since Zak-OTFS does not require subdividing the frame into two halves, we achieve a Doppler resolution of $\nicefrac{1}{T}$, i.e., a $2 \times$ improvement over the FMCW-based architecture from Section~\ref{subsec:prelim_zak}. Moreover, since we achieve polarimetric radar within a single frame interval, our approach has $2 \times$ better latency over the FMCW-based approach.

\subsection{Overview of Zak-OTFS}
\label{subsec:prelim_zak}

The Zak-OTFS carrier waveform is a pulse in the delay-Doppler (DD) domain, formally a quasi-periodic localized function termed the \emph{DD pulsone}\footnote{\textcolor{black}{Termed ``pulsone'' due to its structure of a pulse train modulated by a tone in the time domain, see~\cite[Fig. 2]{bitspaper1} for an illustration.}}. The DD pulsone is characterized by a delay period $\tau_p$ and a Doppler period $\nu_p$, with $\tau_p \nu_p = 1$. The DD pulsone occupies infinite time and bandwidth. For practical implementation, the DD pulsone is limited to a time interval $T$ and a bandwidth $B$ via DD domain pulse shaping. The DD pulsone defines an orthonormal basis within the delay and Doppler periods with $BT = MN$ basis elements at $M = \nicefrac{\tau_p}{\nicefrac{1}{B}} = B\tau_p$ distinct locations along delay and $N = \nicefrac{\nu_p}{\nicefrac{1}{T}} = T\nu_p$ distinct locations along Doppler. %The discrete time domain (TD) representation of a DD pulsone on converting the DD signal to time via the inverse Zak transform followed by Nyquist sampling is~\cite{otfs_book,bitspaper1,bitspaper2}:

\textcolor{black}{The DD pulsone is converted to a time domain (TD) waveform via the inverse Zak transform~\cite{otfs_book,bitspaper1,bitspaper2}. After Nyquist sampling, the discrete TD pulsone waveform is~\cite{otfs_book,bitspaper1,bitspaper2}:}
\begin{align}
    \label{eq:sys_model1}
    \mathbf{p}_{(k_0,l_0)}[n] &= \frac{1}{\sqrt{N}} \sum_{d \in \mathbb{Z}} e^{\frac{j2\pi}{N} d l_0} \delta[n-k_0-dM],
\end{align}
where $k_0 \in \mathbb{Z}_{M}$ indexes the location of the pulsone as a multiple of the delay resolution $\nicefrac{1}{B} = \nicefrac{\tau_p}{M}$, and $l_0 \in \mathbb{Z}_{N}$ indexes the location of the pulsone as a multiple of the Doppler resolution $\nicefrac{1}{T} = \nicefrac{\nu_p}{N}$. The discrete TD signal on mounting $MN$ information symbols on the TD pulsones in~\eqref{eq:sys_model1} is:
\begin{align}
    \label{eq:sys_model2}
    \mathbf{x}[n] &= \sum_{k_0=0}^{M-1} \sum_{l_0=0}^{N-1} \mathbf{X}[k_0,l_0] \mathbf{p}_{(k_0,l_0)}[n],
\end{align}
where $\mathbf{X}$ denotes the $M \times N$ array of information symbols. %It is easily shown~\cite{otfs_book,EURASIP2025} that the TD signal $\mathbf{x}$ is an $MN$-periodic sequence given by the inverse discrete Zak transform (IDZT)~\cite{dzt} of the array $\mathbf{X}$. 

After pulse shaping, the transmitted signal interacts with the scattering environment and is matched filtered at the receiver. The discrete TD received signal is given by~\cite{otfs_book,Mehrotra2025_WCLSpread,EURASIP2025}:
\begin{align}
    \label{eq:sys_model3}
    \mathbf{y}[n]\!&=\!\sum_{k,l \in \mathbb{Z}_{MN}} \mathbf{h}_{\mathsf{eff}}[k,l] \mathbf{x}[(n-k)_{{}_{MN}}] e^{\frac{j2\pi}{MN}l(n-k)}\!+\!\mathbf{w}[n],
\end{align}
where $\mathbf{h}_{\mathsf{eff}}[k,l]$ denotes the \emph{effective channel}\footnote{The effective channel approximates the physical channel when all paths are resolvable in delay with bandwidth $B$ and in Doppler with time $T$.} that encompasses the effects of the physical scattering environment and transmit \& receive pulse shaping/matched filtering \cite[Eq. (7)]{bitspaper2}, and $\mathbf{w}[n]$ denotes the additive noise at the receiver. For a scattering environment with $P$ paths/targets, let $\mathbf{h}_{\mathrm{phy}}(\tau,\nu) = \sum_{t=1}^{P} h^{(p)} \delta(\tau-\tau_t) \delta(\nu-\nu_t)$ denote the corresponding channel representation in the continuous DD domain. The effective channel is given by samples of the continuous effective channel, $\mathbf{h}_{\mathrm{eff}}[k,l] = \mathbf{h}_{\mathrm{eff}}\big(\tau = \frac{k\tau_p}{M},\nu = \frac{l\nu_p}{N}\big)$, where~\cite{otfs_book,bitspaper1,bitspaper2}:
\begin{align}
    \label{eq:heff1}
    \mathbf{h}_{\mathrm{eff}}(\tau,\nu) &= \mathbf{w}_{{}_\mathrm{RX}}(\tau,\nu) *_\sigma \mathbf{h}_{\mathrm{phy}}(\tau,\nu) *_\sigma \mathbf{w}_{{}_\mathrm{TX}}(\tau,\nu).
\end{align}

In~\eqref{eq:heff1}, $\mathbf{w}_{{}_\mathrm{TX}}(\tau,\nu)$ denotes the transmit pulse shaping filter, e.g., $\mathbf{w}_{{}_\mathrm{TX}}(\tau,\nu) = \sqrt{BT}~\mathrm{sinc}(B\tau)~\mathrm{sinc}(T\nu)$ for sinc pulse shaping~\cite{Calderbank2025_isac}, $\mathbf{w}_{{}_\mathrm{RX}}(\tau,\nu) = e^{j2\pi\nu\tau} \mathbf{w}_{{}_\mathrm{TX}}^*(-\tau,-\nu)$ denotes the receiver matched filter, and $*_\sigma$ denotes twisted convolution\footnote{$a(\tau,\nu)*_\sigma b(\tau,\nu) = \iint a(\tau',\nu') b(\tau-\tau',\nu-\nu') e^{j2\pi\nu'(\tau-\tau')} d\tau' d\nu'$.}.

The effective channel is estimated at the receiver via the \emph{cross-ambiguity function}\footnote{When $\mathbf{y} = \mathbf{x}$, the expression $\mathbf{A}_{\mathbf{x},\mathbf{x}}[k,l]$ is called the self-ambiguity.}~\cite{Calderbank2025_isac,zakotfs_ltv,EURASIP2025,Mehrotra2025_WCLSpread}:
\begin{align}
    \label{eq:sys_model4}
    \widehat{\mathbf{h}}_{\mathsf{eff}}[k,l] &= \mathbf{A}_{\mathbf{y},\mathbf{x}}[k,l] \nonumber \\
    &= \sum_{n=0}^{MN-1} \mathbf{y}[n] \mathbf{x}^{*}[(n-k)_{{}_{MN}}]e^{-\frac{j2\pi}{MN}l(n-k)},
\end{align}
whose peaks in the absolute value indicate delay and Doppler bins of potential targets in the scattering environment. It has been shown in~\cite{zakotfs_ltv,EURASIP2025} that computing the cross-ambiguity function in Zak-OTFS requires only $\mathcal{O}(BT \log T)$ complexity.

Accurate channel estimation is possible when the sequence $\mathbf{x}$ satisfies the \emph{crystallization condition}~\cite{otfs_book,bitspaper1,bitspaper2,Calderbank2025_isac,Mehrotra2025_WCLSpread}. Let $\mathcal{S} = \big\{(k,l)\big| \big|\mathbf{A}_{\mathbf{x},\mathbf{x}}[k, l]\big| = 1 \big\}$ denote the DD locations where the self-ambiguity function of $\mathbf{x}$ is unimodular, and let $\mathcal{C}$ denote the maximum DD support of the scattering environment\footnote{e.g., $\mathcal{C} = [k_{\min},k_{\max}] \times [l_{\min},l_{\max}]$ based on prior knowledge of the minimum/maximum delay and Doppler spreads of the scattering environment.}. The crystallization condition requires:
\begin{equation}
    \label{eq:cryst}
    \bigg(\bigcup_{(k,l) \in \mathcal{S}} \big(\mathcal{C} + (k,l)\big) \bigg) \cap \bigg(\bigcup_{(k',l') \in \mathcal{S}} \big(\mathcal{C} + (k',l')\big) \bigg) = \emptyset,
\end{equation}
where $(k,l) \neq (k',l')$. In other words, translates of the channel support by locations where the self-ambiguity function is unimodular must not overlap for accurate channel estimation.

\subsection{Extension to Polarimetry}
\label{subsec:pol_zak}

% We consider a Zak-OTFS-based radar equipped with dual-polarized antennas at the transmitter and receiver. Let $\mathsf{V}, \mathsf{H}$ denote vertical and horizontal polarization as before. The system model in~\eqref{eq:sys_model3} is extended to dual-polarization as:
The system model in~\eqref{eq:sys_model3} is extended to polarimetry as:
\begin{align}
    \label{eq:pol1}
    \mathbf{y}^{(j)}[n]\!&=\!\sum_{i \in \{\mathsf{V},\mathsf{H}\}}\!\sum_{k,l \in \mathbb{Z}_{MN}}\!\mathbf{h}_{\mathsf{eff}}^{(j,i)}[k,l] \mathbf{x}^{(i)}[(n-k)_{{}_{MN}}] e^{\frac{j2\pi}{MN}l(n-k)} \nonumber \\ &~~~~~~~~~~~~~~~~~~~~+ \mathbf{w}^{(j)}[n],~i,j \in \{\mathsf{V},\mathsf{H}\},
\end{align}
where $\mathbf{x}^{(i)}$ denotes the signal transmitted by the $i$-polarized transmit antenna, $\mathbf{h}_{\mathsf{eff}}^{(j, i)}[k, l]$ denotes the effective channel between the $i$-polarized transmit antenna and the $j$-polarized receive antenna, and $\mathbf{w}^{(j)}[n]$ denotes the additive noise at the $j$-polarized receive antenna. In~\eqref{eq:pol1}, the polarimetric effective channel $\mathbf{h}_{\mathsf{eff}}^{(j, i)}[k, l]$ is defined similarly to~\eqref{eq:heff1} using the polarimetric continuous DD channel representation, $\mathbf{h}_{\mathrm{phy}}^{(j, i)}(\tau,\nu) = \sum_{t=1}^{P} h^{(p)}_{ji} \delta(\tau-\tau_t) \delta(\nu-\nu_t)$, where $h^{(p)}_{ji}$ is $(j,i)$th entry of the matrix $\mathbf{H}^{(p)}$ in~\eqref{eq:pol_prelim1}.

%For a scattering environment with $P$ targets, let $\mathbf{H}^{(p)}$ denote the $2 \times 2$ polarimetric scattering matrix of the $t$th target~\cite{howard2007simple}:
% \begin{align}
%     \label{eq:pol_prelim1}
%     \mathbf{H}^{(p)} &= \begin{bmatrix}
%         h^{(p)}_{\mathsf{HH}} & h^{(p)}_{\mathsf{HV}} \\ h^{(p)}_{\mathsf{VH}} & h^{(p)}_{\mathsf{VV}}
%     \end{bmatrix}.
% \end{align}

% \begin{align}
%     \label{eq:pol1}
%     \mathbf{y}^{(j)}[k,l] = \sum_{i \in \{\mathsf{V},\mathsf{H}\}} \mathbf{h}_{\mathsf{eff}}^{(j,i)}[k,l] \ast_{\sigma} \mathbf{x}^{(i)}[k,l] + \tilde{\mathbf{n}}^{(j)}[k, l],
% \end{align}
% where $j \in \{\mathsf{V},\mathsf{H}\}$ and $\mathbf{h}_{\mathsf{eff}}^{(j, i)}[k, l]$ denotes the effective channel between the $i$-polarized transmit antenna to $j$-polarized receive antenna. The assumption here is that both the transmitter and receiver are equipped with dual-polarized antennas. As described in the previous section we consider $\mathbf{x}^{(\mathsf{H})}[k, l]$ to be point pilot with the non-zero entry at ($k_p, l_p$) and $\mathbf{x}^{(\mathsf{V})}[k, l]$ to be a spread pilot.

\subsection{Proposed Approach for Instantaneous Polarimetry}
\label{subsec:prop_appr}

We enable instantaneous polarimetry at \textcolor{black}{near-linear} complexity by designing mutually unbiased sequences via Zak-OTFS. To that end, in the following we define the generalized discrete affine Fourier transform (GDAFT)~\cite{EURASIP2025,Mehrotra2025_WCLSpread}, which maps pulsones in~\eqref{eq:sys_model1} to mutually unbiased spread waveforms. We have utilized the GDAFT to design radar waveform libraries in~\cite{EURASIP2025} and for spread carrier communication in~\cite{Mehrotra2025_WCLSpread}.

\begin{definition}[\cite{EURASIP2025,Mehrotra2025_WCLSpread}]
    \label{def:gdaft}
    The generalized discrete affine Fourier transform (GDAFT) of an $MN$-length sequence $\mathbf{x}$ is:
    \begin{align*}
        \mathcal{F}_a\mathbf{x}[n] = \frac{1}{\sqrt{MN}} \sum_{m=0}^{MN-1} e^{\frac{j2\pi}{MN} (An^2+Bnm+Cm^2)} \mathbf{x}[m],
    \end{align*}
    where $n\in\{0,\cdots,MN-1\}$, $A,B,C$ are co-prime to $MN$.
\end{definition}

\begin{theorem}[\cite{EURASIP2025,Mehrotra2025_WCLSpread}]
    \label{thm:gdaft_pp_tdcazac}
    The GDAFT in Definition~\ref{def:gdaft} maps the discrete time pulsone in~\eqref{eq:sys_model1} localized at $(k_0,l_0)$ in the discrete DD domain to the spread carrier sequence:
    \begin{align*}
        \mathbf{c}[n] = \mathcal{F}_a\mathbf{p}_{(k_0,l_0)}[n] = &\frac{e^{\frac{j2\pi}{MN} (An^2+Bnk_0 +Ck_0^2)}}{\sqrt{MN}} \epsilon_N \left(\frac{CM}{N}\right)_J \nonumber \\ 
        & \times e^{-\frac{j2\pi}{N} (4CM)_N^{-1} (Bn + l_0 + 2Ck_0)^2},
    \end{align*}
    where $\epsilon_{N} = 1$ if $N \equiv 1 \bmod 4$ \& $\epsilon_{N} = j$ if $N \equiv 3 \bmod 4$, and $\left(\frac{a}{b}\right)_J$ denotes the Jacobi symbol.
\end{theorem}

% Two commonly used sequences $\mathbf{x}$ for effective channel estimation are pulsones in~\eqref{eq:sys_model1} and CAZAC (constant amplitude zero autocorrelation) sequences~\cite{Calderbank2025_isac,preamblepaper,EURASIP2025,Mehrotra2025_WCLSpread}:
% \begin{align}
%     \label{eq:cazac}
%     \mathbf{c}[n] &= \frac{1}{\sqrt{MN}} e^{\frac{j2\pi}{MN}(\alpha n^2 + \beta n + \gamma)},
% \end{align}
% where $\alpha, \beta, \gamma \in \mathbb{Z}_{MN}$ with $(\alpha,MN) = 1$. Special cases of~\eqref{eq:cazac} are Zadoff-Chu (ZC) sequences~\cite{zadoff_chu} with root $u$ when $\alpha = \beta = \nicefrac{u}{2}$, $\gamma = 0$ and Wiener sequences~\cite{benedetto_cazac} when $\beta = \gamma = 0$.

A useful consequence of Theorem~\ref{thm:gdaft_pp_tdcazac} is that the output of the GDAFT is mutually unbiased to the pulsone~\cite{EURASIP2025,Mehrotra2025_WCLSpread}:
\begin{align}
    \label{eq:unbiased}
    \mathbf{A}_{\mathbf{c},\mathbf{p}_{(k_0,l_0)}}[k,l] &= \frac{C_{(k_0,l_0)}[k,l]}{\sqrt{MN}},
\end{align}
where $C_{(k_0,l_0)}[k,l]$ is a complex phase, $\big|C_{(k_0,l_0)}[k,l] \big| = 1$, and $\mathbf{A}_{\mathbf{y},\mathbf{x}}[k,l]$ denotes the cross-ambiguity function as in~\eqref{eq:sys_model4}. Moreover, the GDAFT \emph{preserves} the \textcolor{black}{near-linear} computational complexity of cross-ambiguity-based channel estimation~\cite{EURASIP2025}.

% Simply put, each sequence looks like \emph{noise} to the other. This property enables \emph{simultaneously} transmitting the superposition of the two sequences and estimating the effective channel from each sequence separately at the receiver. This property has been utilized for integrating sensing \& communication in the same radio resources in~\cite{Calderbank2025_isac} and for grant-free multiple access in~\cite{preamblepaper}. In the next Section, we describe how to utilize~\eqref{eq:unbiased} to enable single-frame radar polarimetry with Zak-OTFS.

% We propose to exploit the mutual unbiasedness property described in~\eqref{eq:unbiased} to enable estimating all four polarimetric combinations of the effective channel $\mathbf{h}_{\mathsf{eff}}^{(j, i)}[k, l]$, $i,j \in \{\mathsf{V},\mathsf{H}\}$ at the receiver. Specifically, in the same frame we propose to transmit mutually unbiased sequences per polarization, e.g., a pulsone per~\eqref{eq:sys_model1} in $\mathsf{H}$ and a CAZAC per~\eqref{eq:cazac} in $\mathsf{V}$:
For instantaneous polarimetry, we transmit the pulsone and the output of the GDAFT in orthogonal polarizations, e.g.,
\begin{align}
    \label{eq:prop1}
    \mathbf{x}^{(\mathsf{H})}[n] = \mathbf{p}_{(k_0,l_0)}[n],~\mathbf{x}^{(\mathsf{V})}[n] = \mathbf{c}[n].
\end{align}
%where each sequence $\mathbf{x}^{(i)}$ is assumed to satisfy the crystallization condition in~\eqref{eq:cryst} with respect to $\mathbf{h}_{\mathsf{eff}}^{(j, i)}[k, l]$, $i,j \in \{\mathsf{V},\mathsf{H}\}$. 

For accurate channel estimation, the GDAFT parameters $A, B, C$ in Theorem~\ref{thm:gdaft_pp_tdcazac} are chosen such that $\mathbf{c}[n]$ satisfies the crystallization condition in~\eqref{eq:cryst} for all four components of the polarimetric effective channel $\mathbf{h}_{\mathsf{eff}}^{(j, i)}[k, l]$, for all $i,j \in \{\mathsf{V},\mathsf{H}\}$. \textcolor{black}{For a detailed discussion on GDAFT parameter selection, see Fig. 1 and the associated example in~\cite[Section IV-C]{Mehrotra2025_WCLSpread}.}

On obtaining the received signals per~\eqref{eq:pol1}, the receiver computes the cross-ambiguity function per~\eqref{eq:sys_model4} between $\mathbf{y}^{(j)}$ and $\mathbf{x}^{(i)}$ to estimate the effective channel $\mathbf{h}_{\mathsf{eff}}^{(j, i)}[k, l]$, for all $i, j \in \{\mathsf{V}, \mathsf{H}\}$. We now show how mutual unbiasedness per~\eqref{eq:unbiased} enables accurate estimation of all four polarimetric effective channels $\mathbf{h}_{\mathsf{eff}}^{(j, i)}[k, l]$. Without loss of generality, we prove the result for the example considered in~\eqref{eq:prop1}.

The estimate of $\mathbf{h}_{\mathsf{eff}}^{(j, i)}[k, l]$ from~\eqref{eq:sys_model4} and~\eqref{eq:pol1} is given by:
\begin{align}
    \label{eq:prop2}
    \widehat{\mathbf{h}}_{\mathsf{eff}}^{(j, i)}[k,l] &= \mathbf{A}_{\mathbf{y}^{(j)},\mathbf{x}^{(i)}}[k,l] \nonumber \\
    &= \sum_{n=0}^{MN-1} \mathbf{y}^{(j)}[n] \big(\mathbf{x}^{(i)}\big)^{*}[(n-k)_{{}_{MN}}]e^{-\frac{j2\pi}{MN}l(n-k)} \nonumber \\
    &=\!\sum_{i' \in \{\mathsf{V},\mathsf{H}\}}\!\sum_{k',l'} \mathbf{h}_{\mathsf{eff}}^{(j,i')}[k',l']\!\sum_{n=0}^{MN-1}\!\mathbf{x}^{(i')}[(n-k')_{{}_{MN}}] \nonumber \\ &~~~~e^{\frac{j2\pi}{MN}l'(n-k')} \big(\mathbf{x}^{(i)}\big)^{*}[(n-k)_{{}_{MN}}]e^{-\frac{j2\pi}{MN}l(n-k)} \nonumber \\ &~~~~~~~~~~~~~~~~~~~~~~~~~~~~~~+ \mathbf{A}_{\mathbf{w}^{(j)},\mathbf{x}^{(i)}}[k,l].
\end{align}

For additive noise $\mathbf{w}^{(j)}$ uncorrelated with the transmitted signals $\mathbf{x}^{(i)}$, we have $\mathbf{A}_{\mathbf{w}^{(j)},\mathbf{x}^{(i)}}[k,l] = 0$, for all $k, l$. On further making the substitution $n' = (n-k')_{{}_{MN}}$ we obtain:
\begin{align}
    \label{eq:prop3}
    \widehat{\mathbf{h}}_{\mathsf{eff}}^{(j, i)}[k,l] &= \sum_{i' \in \{\mathsf{V},\mathsf{H}\}}\!\sum_{k',l'} \mathbf{h}_{\mathsf{eff}}^{(j,i')}[k',l']\!\sum_{n'=0}^{MN-1}\!\mathbf{x}^{(i')}[n'] e^{\frac{j2\pi}{MN}l'n'} \nonumber \\ &~~\big(\mathbf{x}^{(i)}\big)^{*}[(n'-(k-k')_{{}_{MN}})_{{}_{MN}}]e^{-\frac{j2\pi}{MN}l(n'-(k-k'))} \nonumber \\
    &= \sum_{i' \in \{\mathsf{V},\mathsf{H}\}}\!\sum_{k',l'} \mathbf{h}_{\mathsf{eff}}^{(j,i')}[k',l'] e^{\frac{j2\pi}{MN}l'(k-k')} \nonumber \\ &~~~~~~~~~~\times \mathbf{A}_{\mathbf{x}^{(i')},\mathbf{x}^{(i)}}[(k-k')_{{}_{MN}},(l-l')_{{}_{MN}}] \nonumber \\
    &= \sum_{i' \in \{\mathsf{V},\mathsf{H}\}} \mathbf{h}_{\mathsf{eff}}^{(j,i')}[k,l] \ast_{\sigma} \mathbf{A}_{\mathbf{x}^{(i')},\mathbf{x}^{(i)}}[k,l],
\end{align}
where $\ast_{\sigma}$ denotes discrete twisted convolution~\cite{otfs_book,bitspaper1,bitspaper2,Calderbank2025_isac}.

The expression in~\eqref{eq:prop3} is the sum of two terms:
\begin{align}
    \label{eq:prop4}
    \widehat{\mathbf{h}}_{\mathsf{eff}}^{(j, i)}[k,l] &= \mathbf{h}_{\mathsf{eff}}^{(j,i)}[k,l] \ast_{\sigma} \mathbf{A}_{\mathbf{x}^{(i)},\mathbf{x}^{(i)}}[k,l] \nonumber \\ &+ \mathbf{h}_{\mathsf{eff}}^{(j,\bar{i})}[k,l] \ast_{\sigma} \mathbf{A}_{\mathbf{x}^{(\bar{i})},\mathbf{x}^{(i)}}[k,l],
\end{align}
where $\bar{i}$ denotes a polarization different from $i$ in the set $\{\mathsf{V},\mathsf{H}\}$. Since each sequence $\mathbf{x}^{(i)}$ satisfies the crystallization condition in~\eqref{eq:cryst}, the first term is simply $\mathbf{h}_{\mathsf{eff}}^{(j,i)}[k,l]$. To simplify the second term, we substitute~\eqref{eq:unbiased} to obtain:
\begin{align}
    \label{eq:prop5}
    \widehat{\mathbf{h}}_{\mathsf{eff}}^{(j, i)}[k,l] &= \mathbf{h}_{\mathsf{eff}}^{(j,i)}[k,l] + \mathbf{h}_{\mathsf{eff}}^{(j,\bar{i})}[k,l] \ast_{\sigma} \frac{C[k,l]}{\sqrt{MN}} \nonumber \\
    &\approx \mathbf{h}_{\mathsf{eff}}^{(j,i)}[k,l],
\end{align}
where $C[k,l]$ is a phase term similar to that in~\eqref{eq:unbiased}. Since the second term is the twisted convolution of the effective channel $\mathbf{h}_{\mathsf{eff}}^{(j,\bar{i})}[k,l]$ with a constant amplitude term, it simply raises the noise floor of the channel estimate. Computing each cross-ambiguity term only incurs $\mathcal{O}(BT \log T)$ complexity\footnote{\textcolor{black}{The complexity reduction from quadratic to near-linear is due to the symmetry of the Zak-OTFS carrier waveform. Specifically, the cross-ambiguity function computation with Zak-OTFS carrier waveforms reduces to an FFT (fast Fourier transform) calculation. For more details, see~\cite[Section 4.3]{EURASIP2025}.}}, and the overall complexity remains \textcolor{black}{near-linear} in $BT$.

%% file: results.tex
\section{Numerical Results}
\label{sec:results}

\begin{figure*}
\centering
\begin{subfigure}{0.32\linewidth}
    \includegraphics[width=\textwidth]{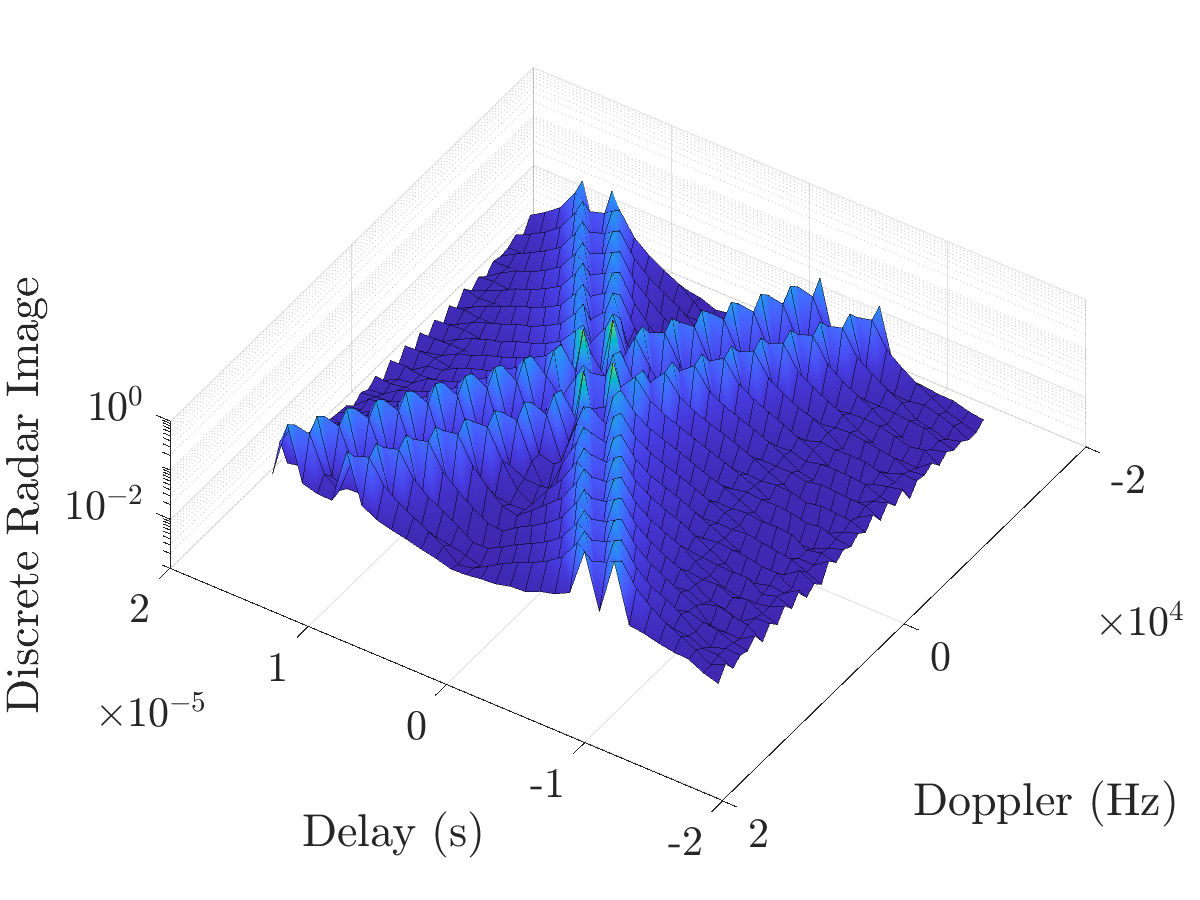}
\caption{FMCW, $\mathsf{HH}$}
    \label{fig:heatmaps_11f}
\end{subfigure}
\begin{subfigure}{0.32\linewidth}
    \includegraphics[width=\textwidth]{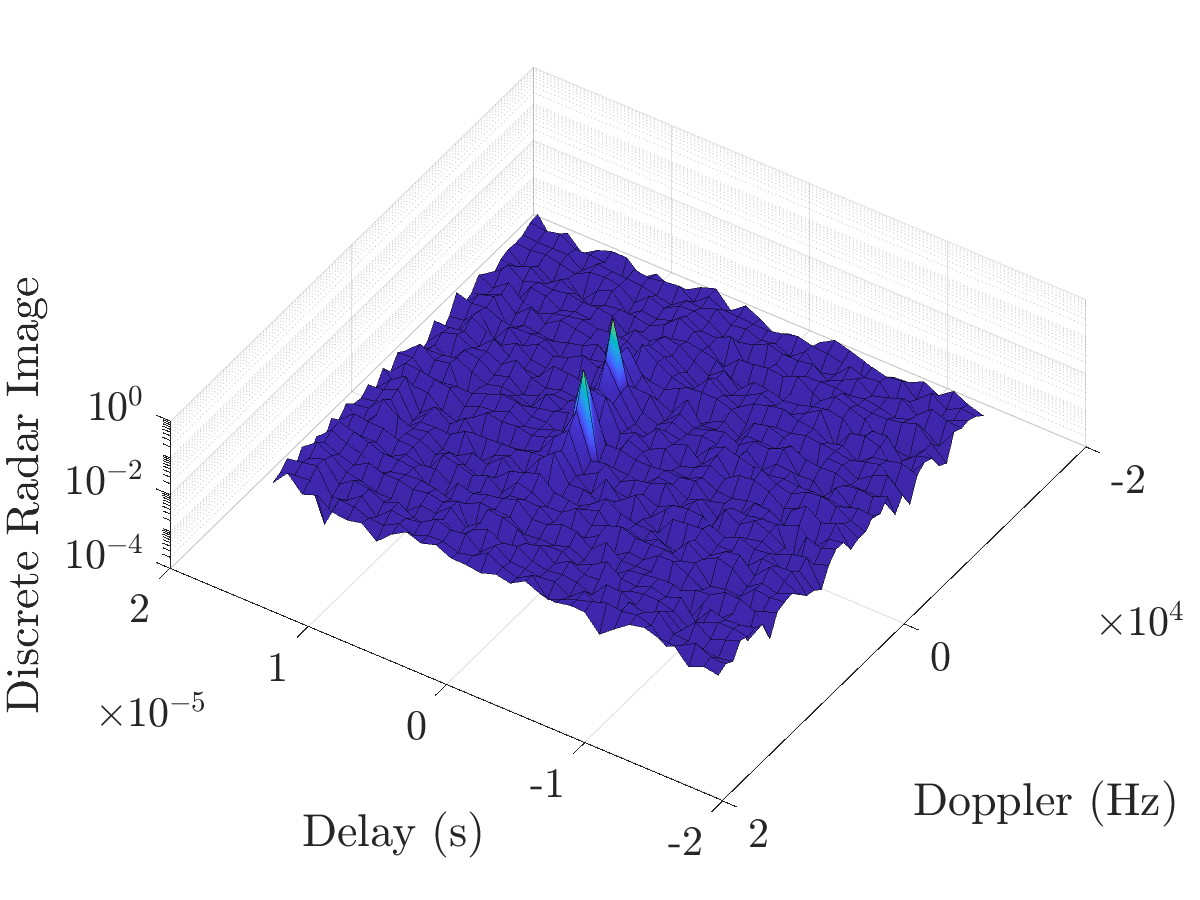}
\caption{Phase-coded, $\mathsf{HH}$}
    \label{fig:heatmaps_11p}
\end{subfigure}
\begin{subfigure}{0.32\linewidth}
    \includegraphics[width=\textwidth]{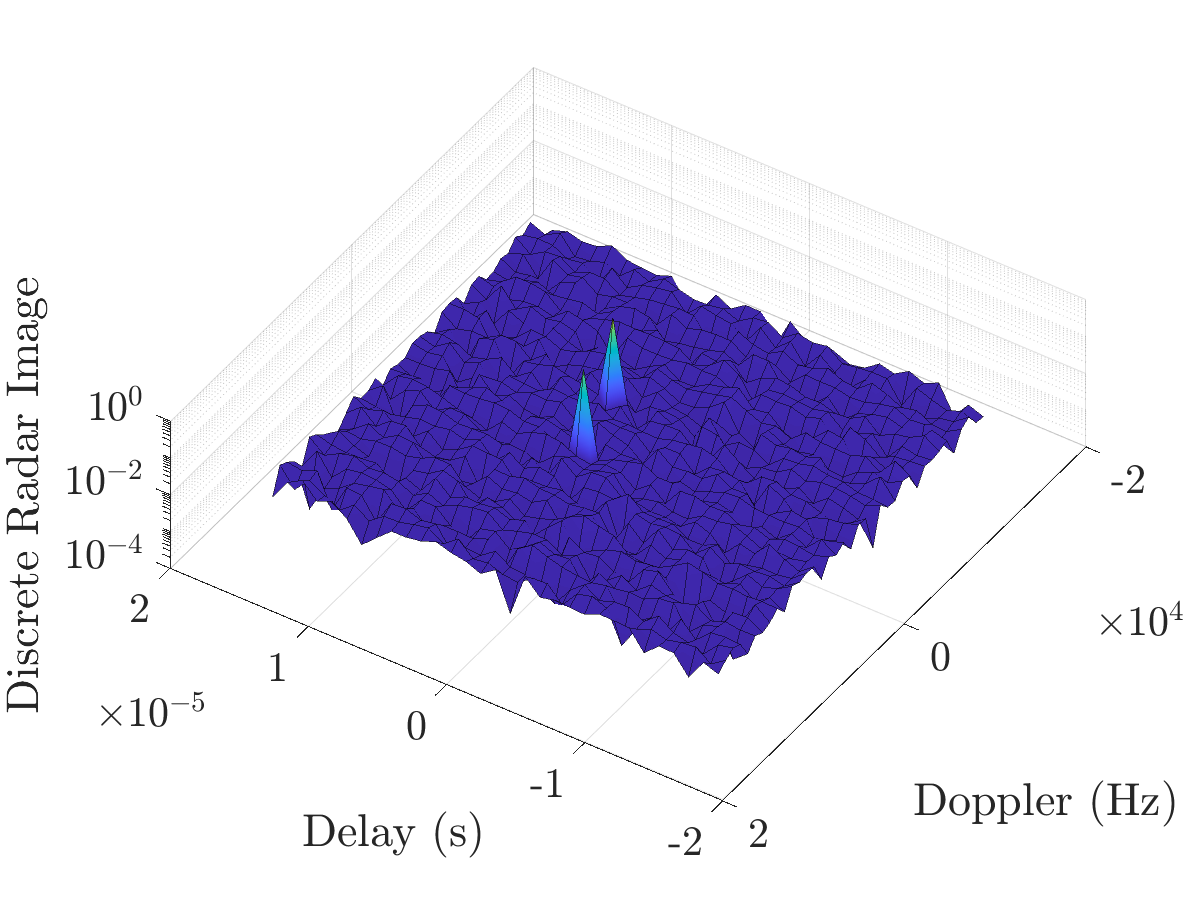}
\caption{Zak-OTFS, $\mathsf{HH}$}
    \label{fig:heatmaps_11z}
\end{subfigure}

\vspace{0.5em}

\begin{subfigure}{0.32\linewidth}
    \includegraphics[width=\textwidth]{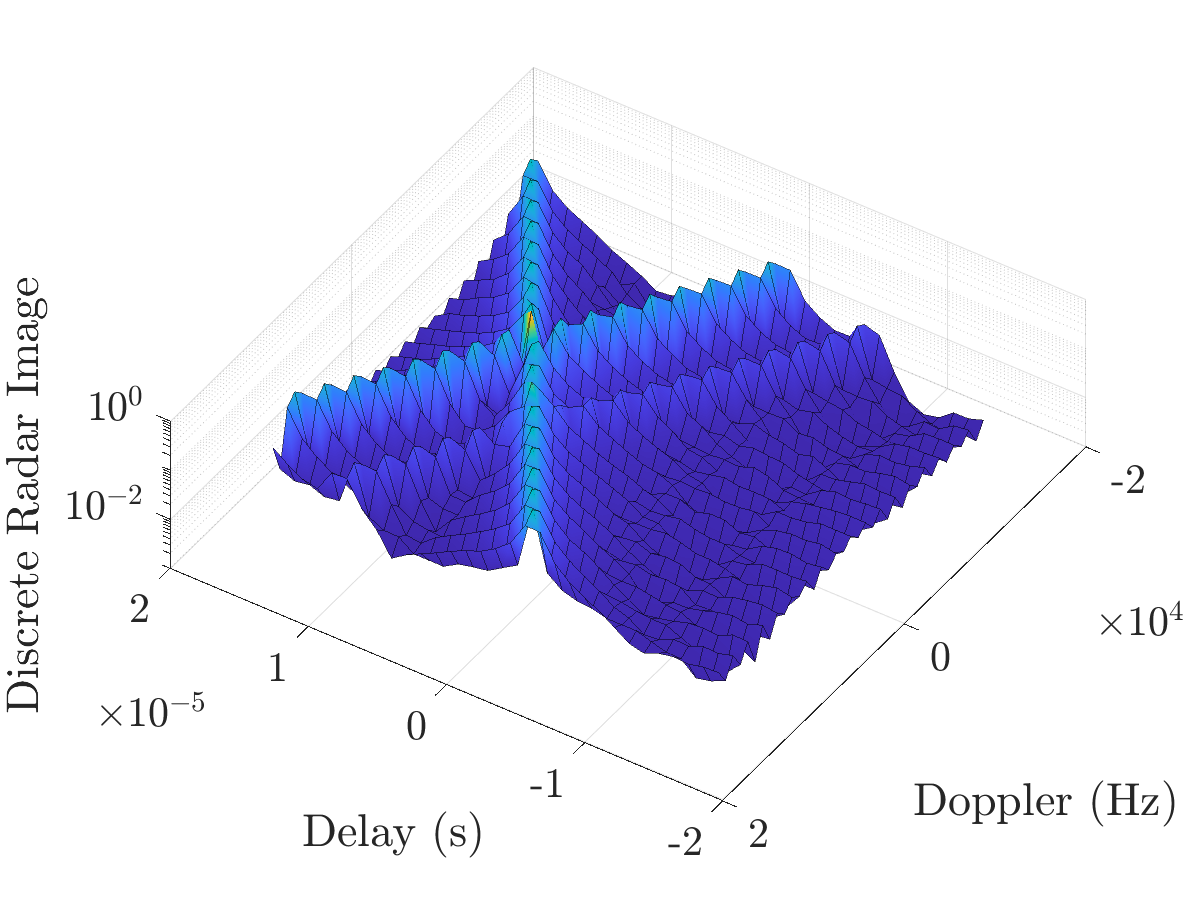}
\caption{FMCW, $\mathsf{VH}$}
    \label{fig:heatmaps_21f}
\end{subfigure}
\begin{subfigure}{0.32\linewidth}
    \includegraphics[width=\textwidth]{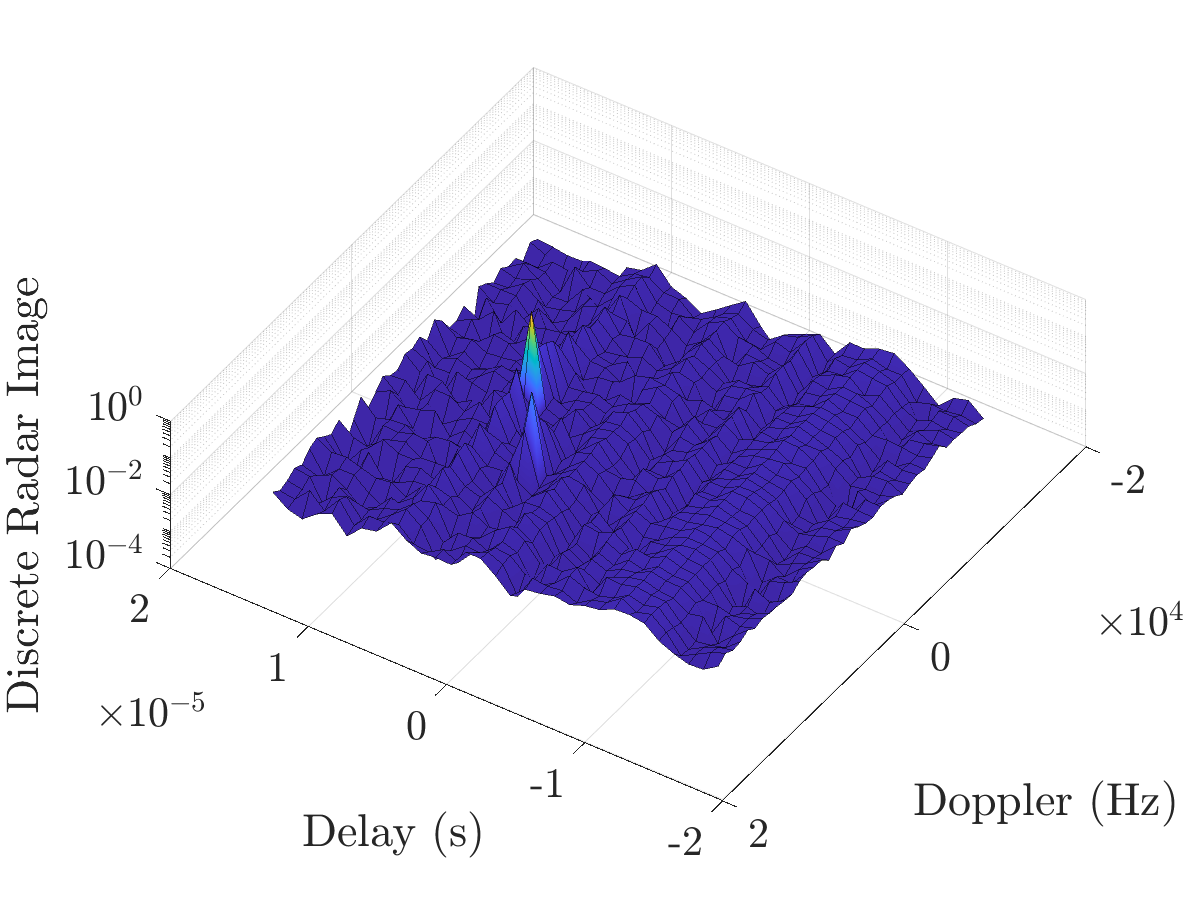}
\caption{Phase-coded, $\mathsf{VH}$}
    \label{fig:heatmaps_21p}
\end{subfigure}
\begin{subfigure}{0.32\linewidth}
    \includegraphics[width=\textwidth]{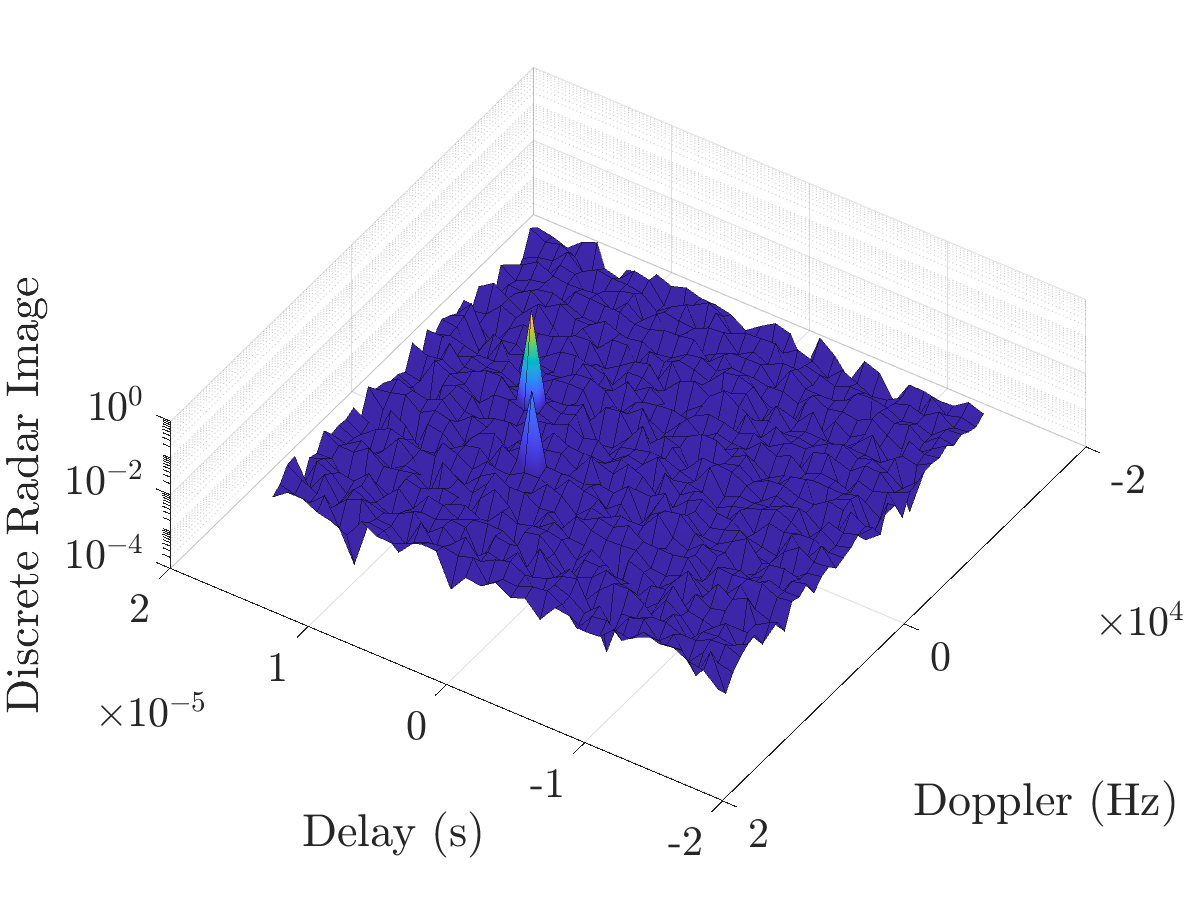}
\caption{Zak-OTFS, $\mathsf{VH}$}
    \label{fig:heatmaps_21z}
\end{subfigure}
\caption{Heatmaps of estimated channels for a four-target environment with two targets with equal $h^{(p)}_{\mathsf{HH}} = 0.7$ \& $h^{(p)}_{\mathsf{HV}} = h^{(p)}_{\mathsf{VH}} = h^{(p)}_{\mathsf{VV}} = 0$, and two targets with unequal $h^{(p)}_{\mathsf{HV}} = h^{(p)}_{\mathsf{VH}} \in \{0.3,0.95\}$ \& $h^{(p)}_{\mathsf{HH}} = h^{(p)}_{\mathsf{VV}} = 0$. (a) \& (d): Sequential polarimetry via FMCW detects two false targets (``ghost targets'') in addition to the two true targets in the $\mathsf{HH}$ channel, and fails to detect the lower energy target in the $\mathsf{VH}$ channel due to the high sidelobes of the waveform. (b)-(c) \& (e)-(f): Instantaneous polarimetry via mutually unbiased phase-coded and Zak-OTFS waveforms detects all four targets correctly in the $\mathsf{HH}$ and $\mathsf{VH}$ channels.} %(c) \& (f): Instantaneous polarimetry via mutually unbiased Zak-OTFS waveforms (Fig.~\ref{fig:pol_block_dia}(\subref{fig:pol_zak})) detects all four targets correctly in the $\mathsf{HH}$ and $\mathsf{VH}$ channels.} %, with small sidelobes due to the sub-optimal choice of the carrier %, with minimal sidelobes due to the optimal choice of the waveform
    \label{fig:heatmaps}
\end{figure*}

We now qualitatively and quantitatively compare the performance of the proposed approach from Section~\ref{sec:pol_zak} with sequential polarimetry via FMCW (Section~\ref{subsec:prelim_pol_fmcw}) and instantaneous polarimetry via phase-coding (Section~\ref{subsec:prelim_pol_phasecoding}). We also compare against uni-polarized systems. \textcolor{black}{Our initial results are limited to target detection and estimation in the presence of noise; extensions to clutter are briefly pursued in Section~\ref{subsec:sim_clutter}.} %All our results in this Section are limited to the detection of a single target ($P = 1$). polarimetric target detection and parameter estimation As mentioned in the Introduction, 

We simulate a monostatic polarimetric radar with frame transmissions of bandwidth $B = 930$ kHz and time $T = 1.2$ ms. For FMCW transmissions, in each frame we simulate up-chirps and down-chirps occupying bandwidth $B$ and time $\nicefrac{T}{2}$ each as described in Section~\ref{subsec:prelim_pol_fmcw} sampled at $f_s = 2B$. For phase-coded transmissions, we consider a rectangular carrier waveform with $BT$ chips of length $\nicefrac{1}{B}$ sampled at $f_s = 2B$, which is modulated by Zadoff-Chu sequences of roots $u \in \{101,107\}$. For Zak-OTFS, we consider a delay period of $\tau_p = 33.33~\mu$s and a Doppler period of $\nu_p = 30$ kHz, which correspond to $M = 31$ and $N = 37$ resolvable locations along delay and Doppler respectively. We assume sinc pulse shaping~\cite{Calderbank2025_isac}. \textcolor{black}{Unless noted otherwise,} we add white Gaussian noise $\mathbf{w}^{(j)}$ to the received signals in~\eqref{eq:pol1}. %with optimal parameters corresponding to $99\%$ energy concentration in bandwidth $B$ and time $T$ 

\subsection{Polarimetric Channel Estimation (Qualitative)}
\label{subsec:sim_heatmaps}

Fig.~\ref{fig:heatmaps} qualitatively compares the performance of polarimetric channel estimation via the cross-ambiguity function\footnote{\textcolor{black}{For our implementation on a CPU cluster with $36$ cores and $64$ GB memory, the median time for cross-ambiguity computation with frame size $M = 31$, $N = 37$ is $2.4363$s for FMCW with median absolute deviation $0.081$s, $2.3445$s for phase-coded with median absolute deviation $0.0868$s, and $0.0012$s for Zak-OTFS with median absolute deviation $45.044 \mu$s, consistent with the computational complexity analysis in Section~\ref{sec:pol_zak}.}} in~\eqref{eq:sys_model4} for a four-target scattering environment. We simulate two targets with equal $h^{(p)}_{\mathsf{HH}} = 0.7$ \& $h^{(p)}_{\mathsf{HV}} = h^{(p)}_{\mathsf{VH}} = h^{(p)}_{\mathsf{VV}} = 0$, and two targets with unequal $h^{(p)}_{\mathsf{HV}} = h^{(p)}_{\mathsf{VH}} \in \{0.3,0.95\}$ \& $h^{(p)}_{\mathsf{HH}} = h^{(p)}_{\mathsf{VV}} = 0$. We observe that polarimetry via FMCW detects false targets in the $\mathsf{HH}$ channel and fails to detect the low energy target in the $\mathsf{VH}$ channel due to high sidelobes of the waveform. In contrast, polarimetry via phase-coding and Zak-OTFS achieves ideal target detection with minimal sidelobes around the target locations.

\subsection{Target Detection \& Parameter Estimation (Quantitative)}
\label{subsec:sim_roc}

\begin{figure*}
\centering
\begin{subfigure}{0.32\linewidth}
    \includegraphics[width=\textwidth]{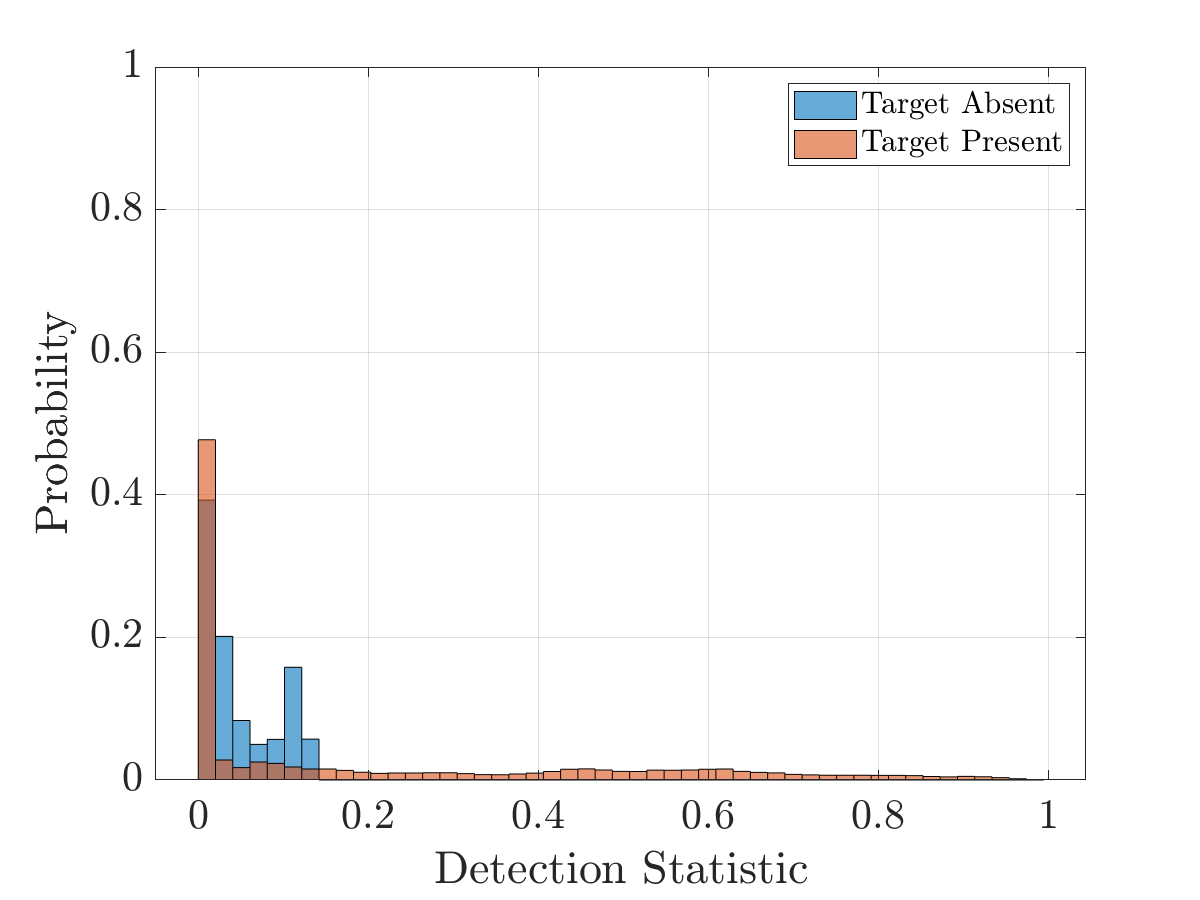}
\caption{FMCW, uni-polarized}
    \label{fig:histograms_fs}
\end{subfigure}
\begin{subfigure}{0.32\linewidth}
    \includegraphics[width=\textwidth]{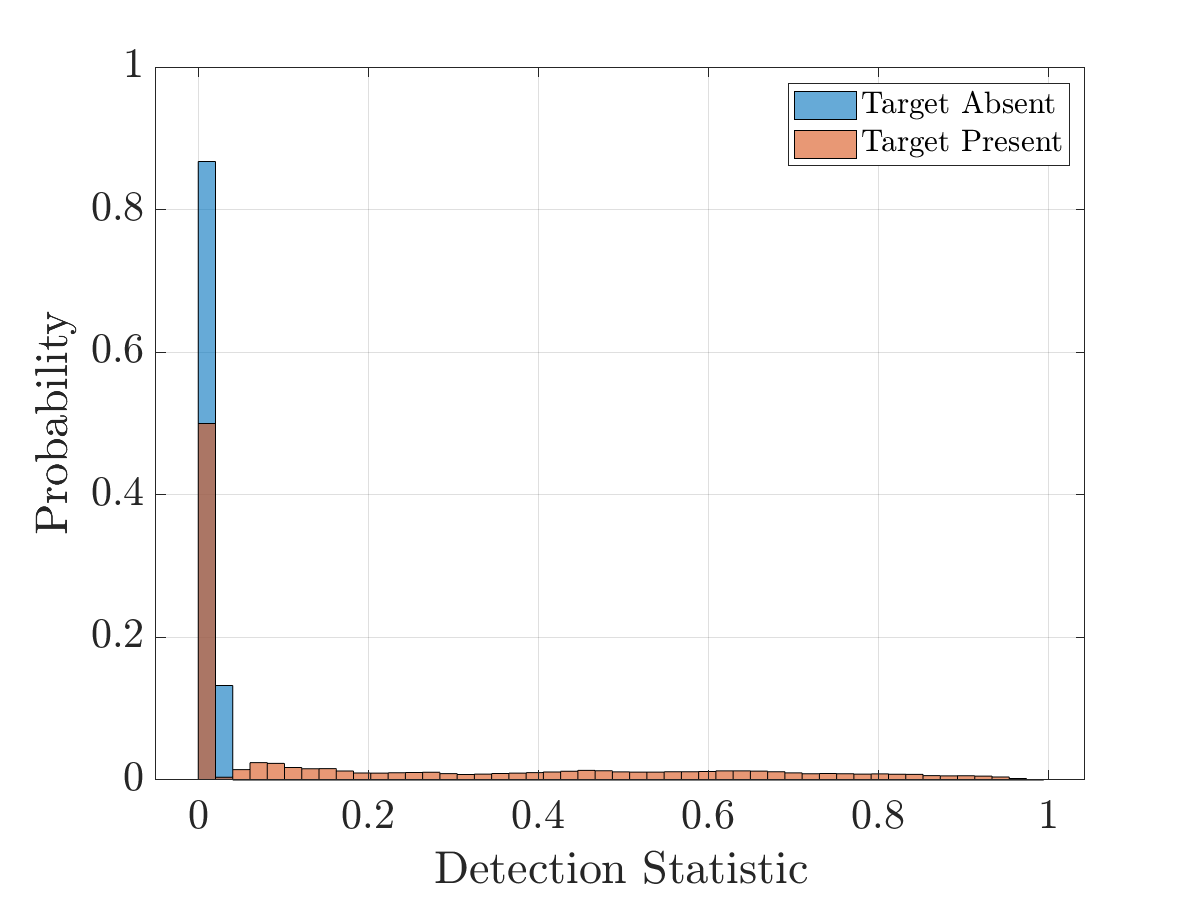}
\caption{Phase-coded, uni-polarized}
    \label{fig:histograms_ps}
\end{subfigure}
\begin{subfigure}{0.32\linewidth}
    \includegraphics[width=\textwidth]{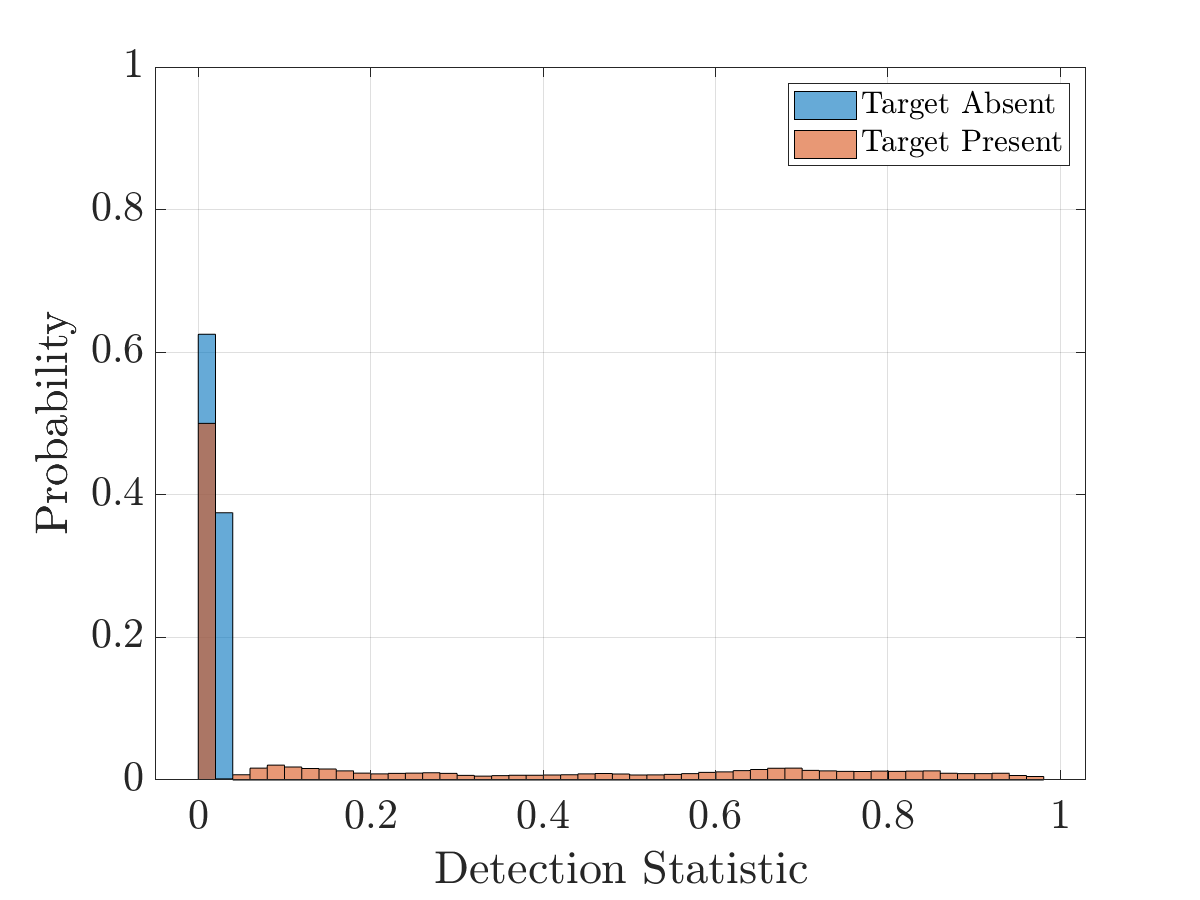}
\caption{\textcolor{black}{Zak-OTFS, uni-polarized}}
    \label{fig:histograms_zs}
\end{subfigure}

\vspace{0.5em}

\begin{subfigure}{0.32\linewidth}
    \includegraphics[width=\textwidth]{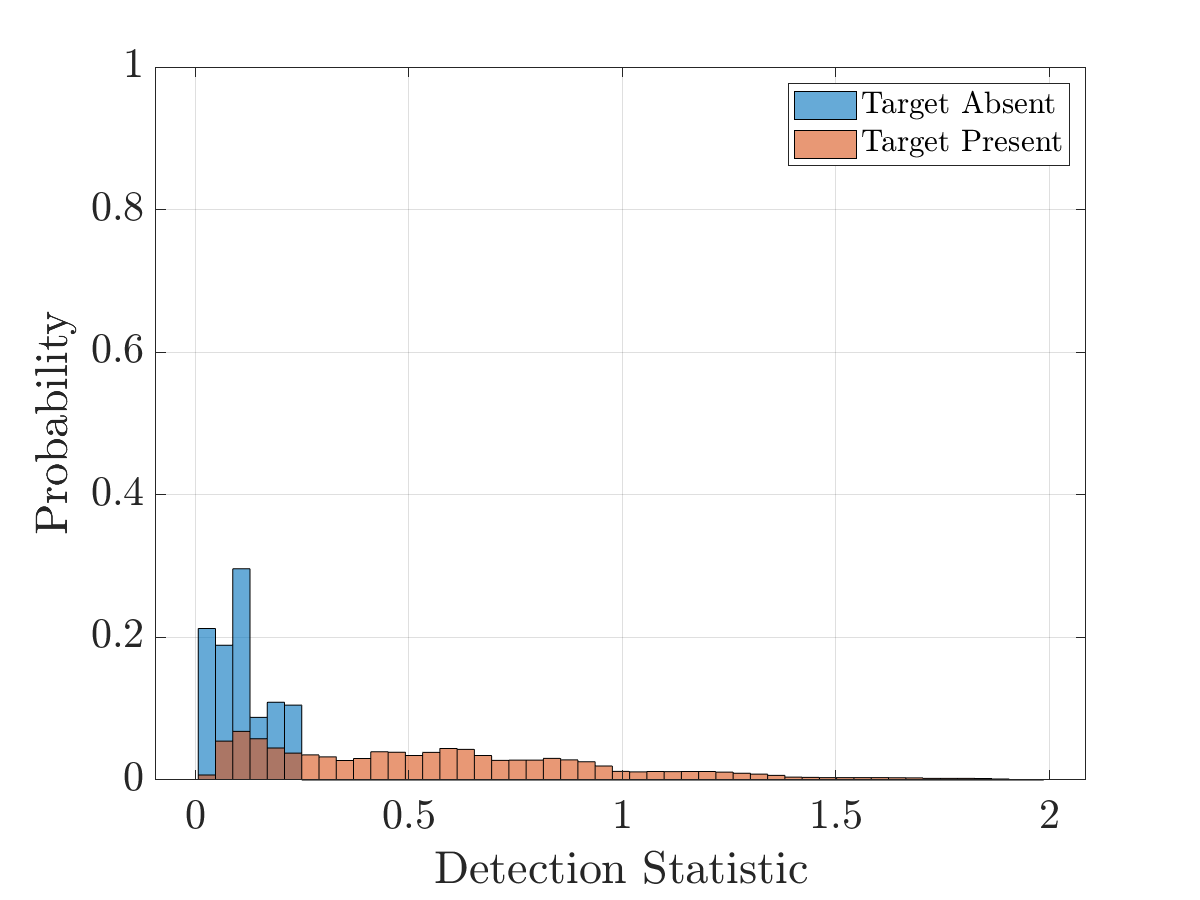}
\caption{FMCW, dual-polarized}
    \label{fig:histograms_fall}
\end{subfigure}
\begin{subfigure}{0.32\linewidth}
    \includegraphics[width=\textwidth]{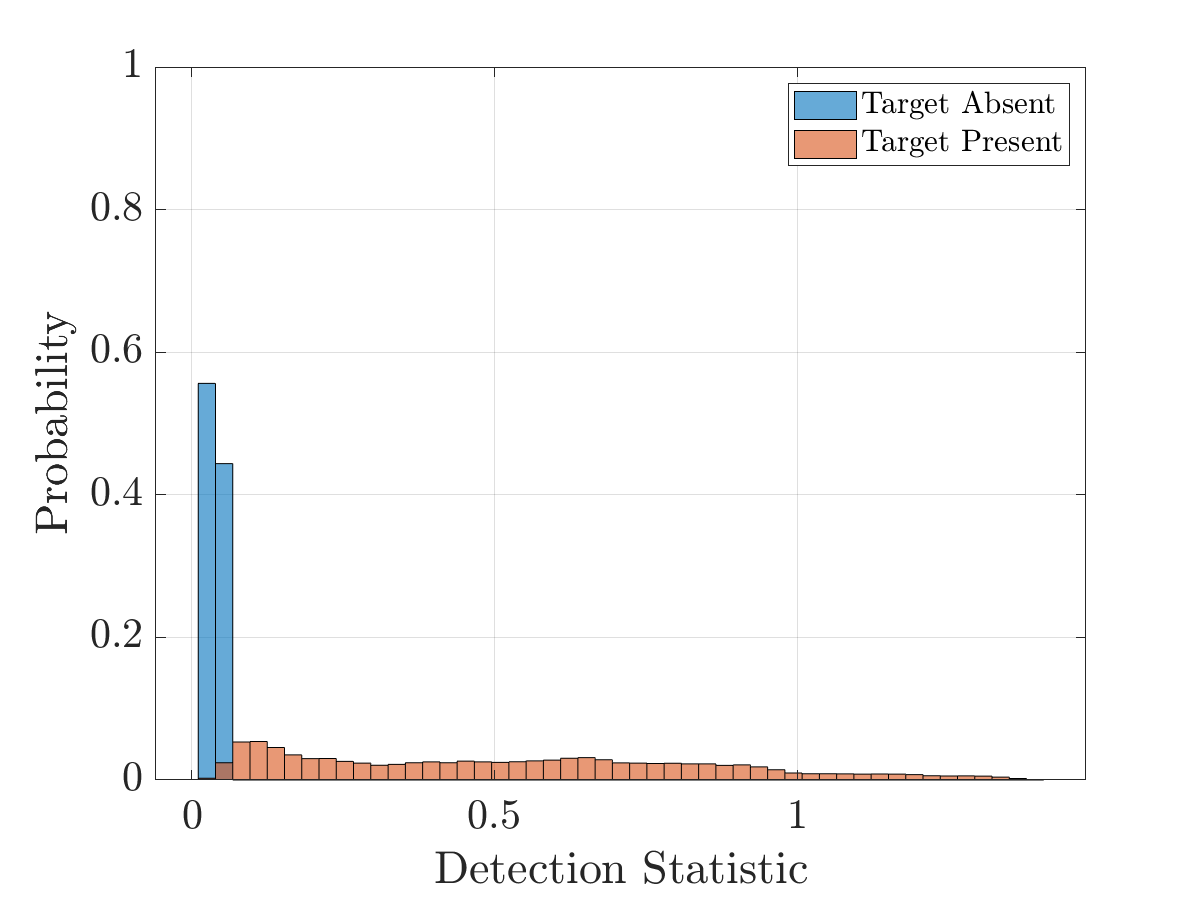}
\caption{Phase-coded, dual-polarized}
    \label{fig:histograms_pall}
\end{subfigure}
\begin{subfigure}{0.32\linewidth}
    \includegraphics[width=\textwidth]{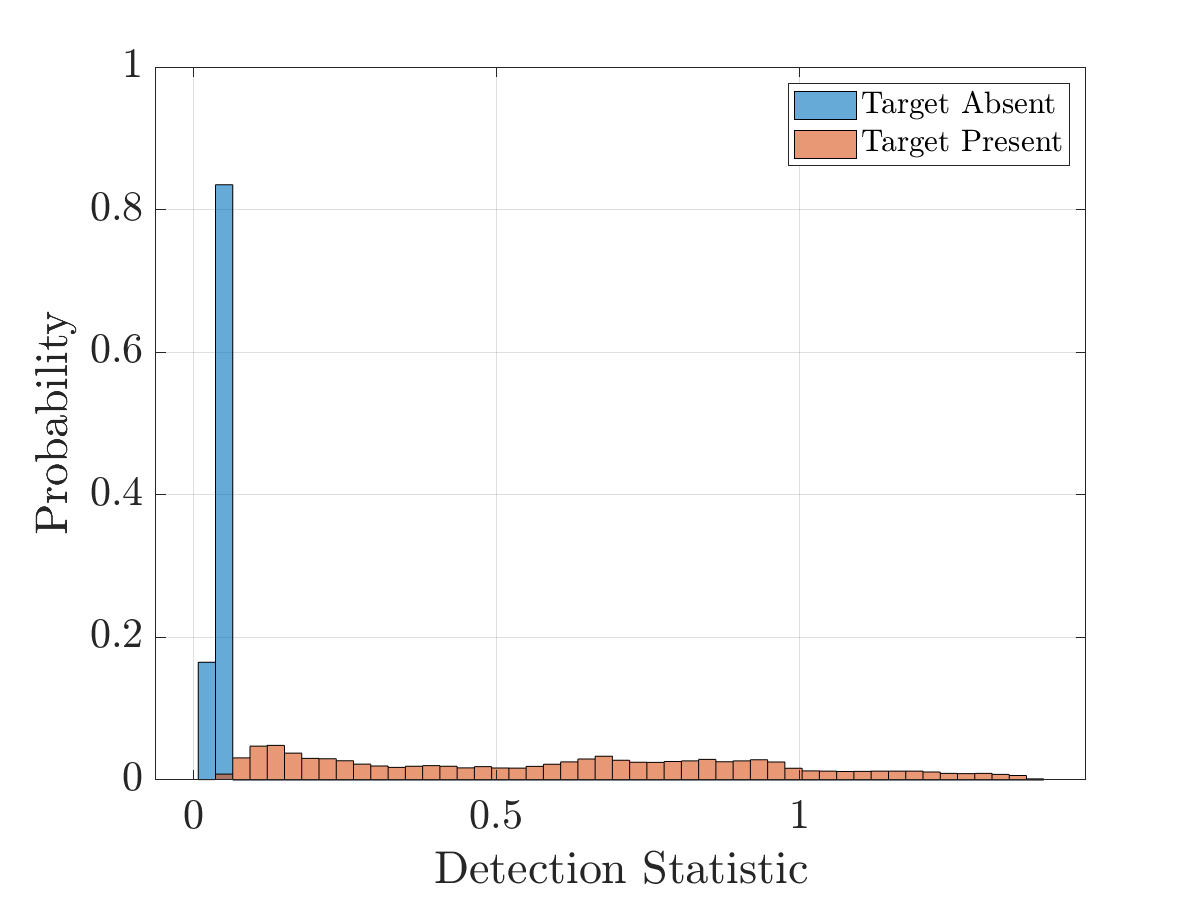}
\caption{Zak-OTFS, dual-polarized}
    \label{fig:histograms_zall}
\end{subfigure}

\caption{Histograms for single polarimetric target detection under the target present and target absent hypotheses. (a)-(c): Uni-polarization is insufficient for detecting polarimetric targets. (d): Dual-polarized FMCW is not optimal for polarimetric target detection due to high waveform sidelobes \& false target detections (cf. Figs.~\ref{fig:heatmaps}(\subref{fig:heatmaps_11f}) \&~\ref{fig:heatmaps}(\subref{fig:heatmaps_21f})). (e)-(f): Dual-polarized phase-coded and Zak-OTFS waveforms are optimal for polarimetric target detection (cf. Figs.~\ref{fig:heatmaps}(\subref{fig:heatmaps_11p})-(\subref{fig:heatmaps_11z}) \&~\ref{fig:heatmaps}(\subref{fig:heatmaps_21p})-(\subref{fig:heatmaps_21z})).}
    \label{fig:histograms}
\end{figure*}

\begin{figure*}
    \centering
    \begin{subfigure}{0.32\linewidth}
    \includegraphics[width=\textwidth]{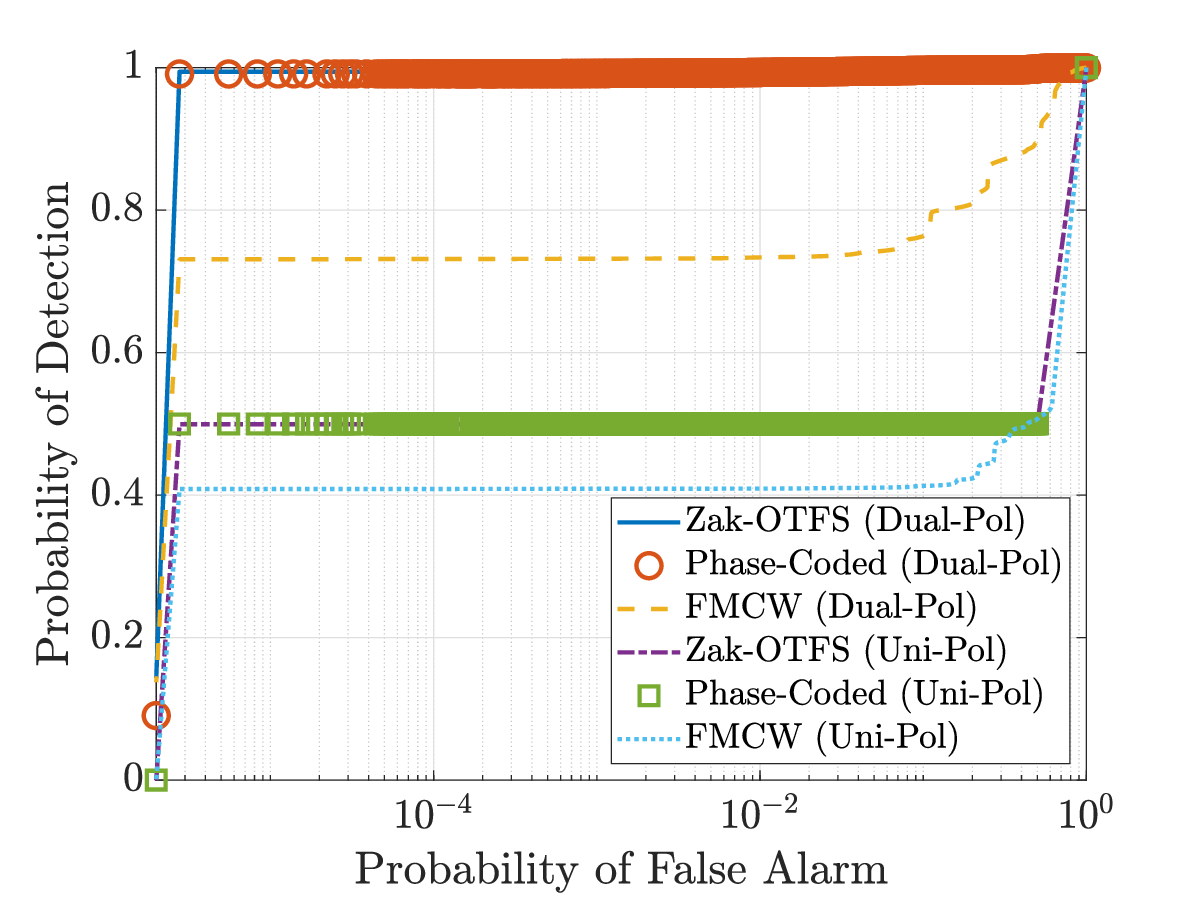}
    \caption{Detection ROC.}
        \label{fig:perf_roc}
    \end{subfigure}
    \begin{subfigure}{0.32\linewidth}
        \includegraphics[width=\textwidth]{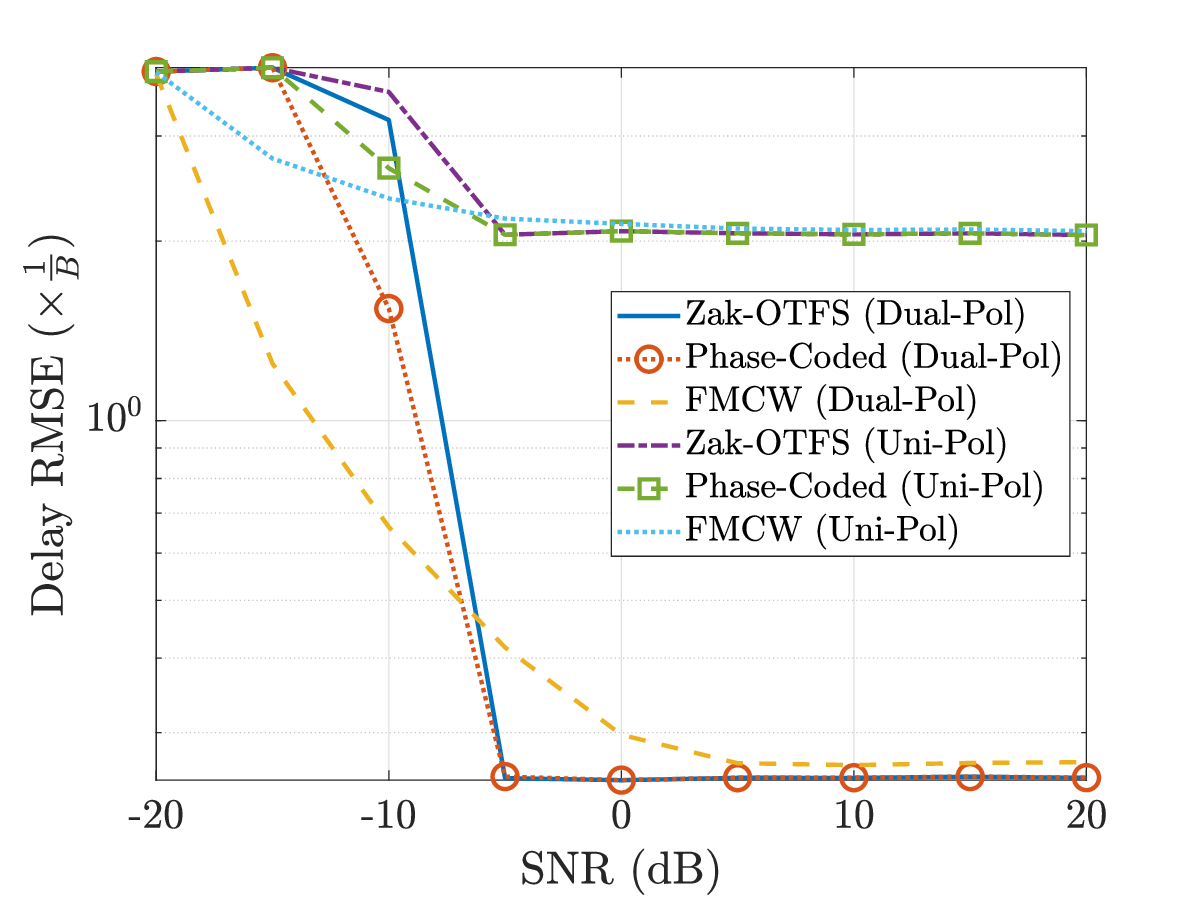}
    \caption{Delay estimation RMSE.}
        \label{fig:perf_rmse_del}
    \end{subfigure}
    \begin{subfigure}{0.32\linewidth}
        \includegraphics[width=\textwidth]{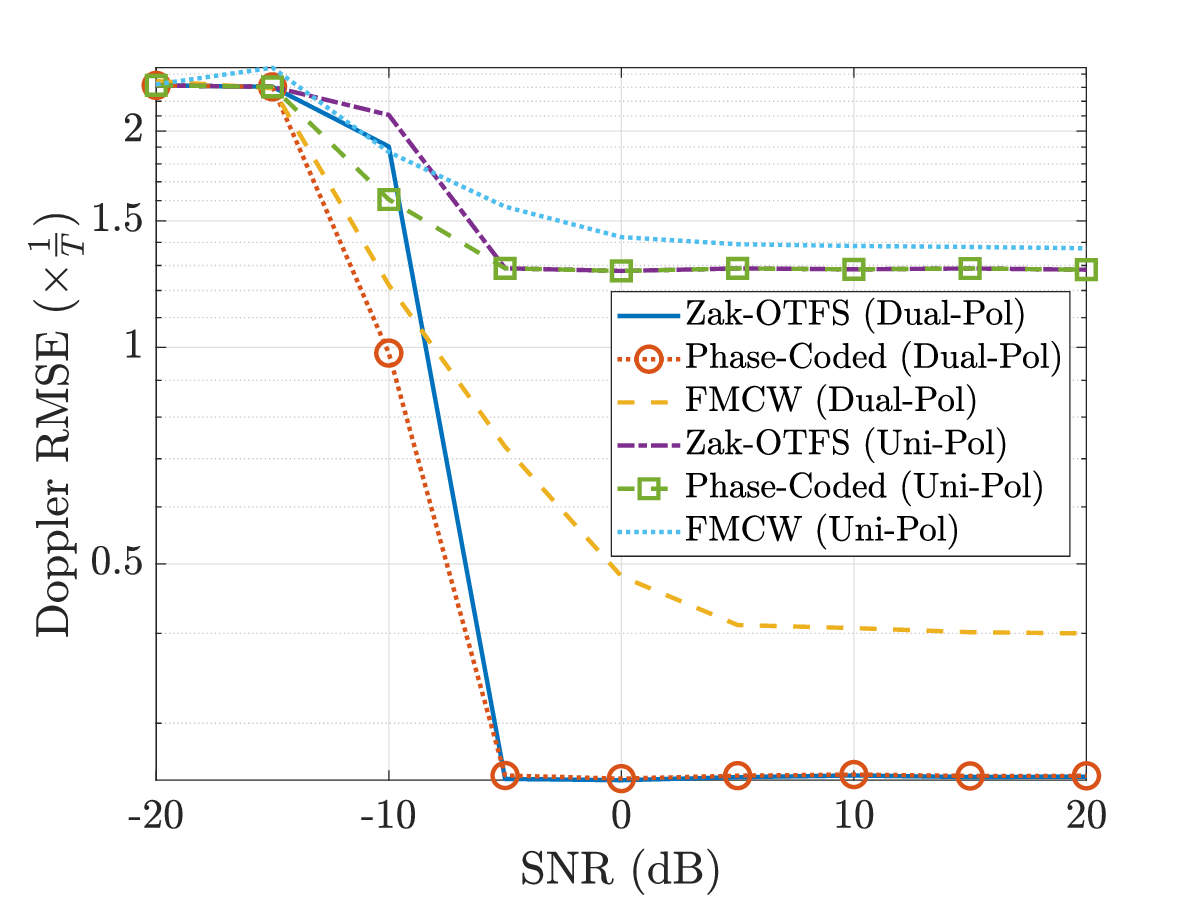}
    \caption{Doppler estimation RMSE.}
        \label{fig:perf_rmse_dop}
    \end{subfigure}
    \caption{Single target detection and estimation performance. (a): Receiver operating characteristic (ROC) curve showing ideal target detection with dual-polarized Zak-OTFS and phase-coded waveforms. The performance degrades with FMCW and/or uni-polarized waveforms. (b)-(c): Root mean squared error (RMSE) for delay and Doppler estimation, normalized by the corresponding delay and Doppler resolutions of $\nicefrac{1}{B}$ and $\nicefrac{1}{T}$. (b): Delay RMSE is similar for dual-polarized Zak-OTFS, phase-coded and FMCW systems at high signal-to-noise ratio (SNR), with significant improvements over uni-polarized waveforms. (c): Doppler RMSE is similar for dual-polarized Zak-OTFS and phase-coded waveforms at high SNR, with $\sim 1.5 \times$ improvement over FMCW due to no loss in Doppler resolution. Significant improvements with dual-polarized vs uni-polarized waveforms.}
    \label{fig:roc}
\end{figure*}

We now quantify the performance of target detection and parameter (delay \& Doppler) estimation. We consider a single target and model the entries of the $2 \times 2$ matrix $\mathbf{H}^{(p)}$ as $h^{(p)}_{\mathsf{HH}} = a \sigma e^{j\phi}$, $h^{(p)}_{\mathsf{HV}} = h^{(p)}_{\mathsf{VH}} = a \sqrt{1-\sigma^2} e^{j\delta}$, $h^{(p)}_{\mathsf{VV}} = b \sigma e^{j\gamma}$, where $a, b \sim \mathsf{Bernoulli}(0.5)$ are i.i.d. symmetric Bernoulli random variables, $\sigma \sim U(0,1)$ is a standard uniform random variable, and $\phi, \delta, \gamma \sim U(0,2\pi)$ are i.i.d. uniform random variables in $[0,2\pi)$. The delay \& Doppler of the target are drawn uniformly at random according to $\tau_t \sim U(0,\nicefrac{\tau_p}{4})$, $\nu_t \sim U(-\nicefrac{\nu_p}{8},\nicefrac{\nu_p}{8})$. We generate $4 \times 10^4$ Monte Carlo instances with signal-to-noise ratio (SNR) ranging from $-20$ dB to $20$ dB. %in steps of $5$ dB. %We compare the performance of uni-polarized and dual-polarized systems based on both FMCW and Zak-OTFS. The uni-polarized systems correspond to transmitting and receiving at the same polarization (either vertical $\mathsf{V}$ or horizontal $\mathsf{H}$). For all systems, we estimate the polarimetric channels via the cross-ambiguity.

\subsubsection{Target Detection}
\label{subsubsec:sim_roc_det}

As the detection criteria, we compare the absolute value of the estimated channel at the delay-Doppler bin corresponding to the target's location (representing the target present hypothesis) with the root-mean-squared value of the channel values at all locations other than the target's location (representing the target absent hypothesis). 

Fig.~\ref{fig:histograms} plots the histograms corresponding to the two hypotheses for all considered systems. Overlapping histograms indicates poor target detectability. The histograms overlap significantly in the uni-polarized systems (Figs.~\ref{fig:histograms}(\subref{fig:histograms_fs})-(\subref{fig:histograms_zs})), since a single polarization is insufficient for estimating the full polarimetric scattering response. Polarimetry via FMCW also has significantly overlapping histograms (Fig.~\ref{fig:histograms}(\subref{fig:histograms_fall})) due to false detections and high sidelobes of the waveform. Polarimetry via phase-coding and Zak-OTFS has minimal overlap between the histograms (Figs.~\ref{fig:histograms}(\subref{fig:histograms_pall})-(\subref{fig:histograms_zall})), indicating their optimality.

Fig.~\ref{fig:roc}(\subref{fig:perf_roc}) plots the receiver operating characteristic (ROC) curve for all considered systems. Consistent with the findings from Fig.~\ref{fig:histograms}, we observe that polarimetry with phase-coded and Zak-OTFS waveforms achieves ideal target detection performance, with performance significantly degrading with FMCW and/or uni-polarized transmissions.

\subsubsection{Parameter Estimation}
\label{subsubsec:sim_roc_est}

For parameter estimation, we first detect peaks in the estimated channel following the procedure outlined in~\cite{Calderbank2025_isac}, which closely mimics the operation of a 2D constant false alarm rate detector from radar signal processing~\cite{Skolnik1980}. We threshold the energy of each channel location by the mean noise energy outside the region of interest $\mathsf{ROI} = [-\Delta \tau,\nicefrac{\tau_p}{4} + \Delta \tau] \times [-\nicefrac{\nu_p}{8} - \Delta \nu,\nicefrac{\nu_p}{8} + \Delta \nu]$ ($\Delta \tau$ and $\Delta \nu$ are guard widths to account for spread due to pulse shaping), scaled by an appropriate factor to achieve $10^{-6}$ false alarm rate~\cite{Skolnik1980}. After thresholding, the delay-Doppler location with the maximum channel energy is the estimated target location. 

To fuse the target parameter estimates obtained across multiple polarimetric components, we compute their weighted average using \emph{entropy-based weights}, $w^{(j,i)} = 1-\nicefrac{\mathcal{H}\big(\widehat{\mathbf{h}}_{\mathsf{eff}}^{(j, i)}\big)}{\log_{2}(MN)}$ for all $i,j \in \{\mathsf{V},\mathsf{H}\}$, where $\mathcal{H}(\mathbf{h}) = -\sum_{k,l} \big(\nicefrac{|\mathbf{h}[k,l]|^{2}}{\sum_{k',l'} |\mathbf{h}[k',l']|^{2}}\big) \log_{2}\big(\nicefrac{|\mathbf{h}[k,l]|^{2}}{\sum_{k',l'} |\mathbf{h}[k',l']|^{2}}\big)$ denotes the entropy~\cite{Li2013_sarentropy} of a DD channel $\mathbf{h}$. Intuitively, such entropy-based weighting prioritizes parameter estimates obtained from polarimetric components with large variation in channel amplitudes (indicating the presence of targets) as opposed to those with little variation in channel amplitudes (indicating noise and the absence of any target). For illustration, consider the extreme case with no targets. Due to mutual unbiasedness, the estimated channel per~\eqref{eq:prop5} in this case has a constant energy level of $\nicefrac{1}{MN}$ and an entropy of $\log_{2}(MN)$, for which the chosen weight is $w^{(j,i)} = 0$, i.e., such components are \emph{not} allowed to bias our estimates.
% we \emph{do not} allow parameter estimates obtained from such polarimetric components to bias our estimates.

Figs.~\ref{fig:roc}(\subref{fig:perf_rmse_del})-(\subref{fig:perf_rmse_dop}) plot the root-mean-squared error (RMSE) for delay and Doppler estimation for all considered systems. The RMSE for delay estimation is similar for polarimetry with FMCW, phase-coding and Zak-OTFS at high SNRs. The Doppler RMSE matches for polarimetry with phase-coding and Zak-OTFS, with $\sim 1.5 \times$ improvement over FMCW, consistent with the explanation in Section~\ref{subsec:prelim_pol_fmcw}. Uni-polarized systems have significantly poorer delay and Doppler RMSE. Note that the RMSEs do not improve beyond a certain threshold due to the inherent resolution limits of the chosen waveforms. Future work will design optimal approaches achieving theoretical bounds, e.g., the Cram\'er-Rao bound~\cite{Vantrees2001}.

%Designing optimal parameter estimation approaches that achieve information theoretic lower bounds, e.g., the Cram\'er-Rao bound~\cite{Vantrees2001}, is a subject of future work. % and the parameter estimation procedure described previously

% the dual-polarized Zak-OTFS system offers the best target detection performance with probability of detection being $0.99$ for all probabilities of false alarm greater than $2.77 \times 10^{-6}$. The target detection is significantly degraded for the dual-polarized FMCW system with probability of detection being $0.99$ only at probabilities of false alarm greater than $0.67$. The uni-polarized FMCW and Zak-OTFS systems do not offer good detection performance, with probability of detection close to $0.5$ across nearly the entire range of probabilities of false alarm. The uni-polarized performance remains slightly better with Zak-OTFS compared to FMCW due to the lower sidelobes with Zak-OTFS.

\subsection{\textcolor{black}{Target Detection in Constant-$\gamma$ Clutter}}
\label{subsec:sim_clutter}

\begin{figure}
    \centering
    \includegraphics[width=\linewidth]{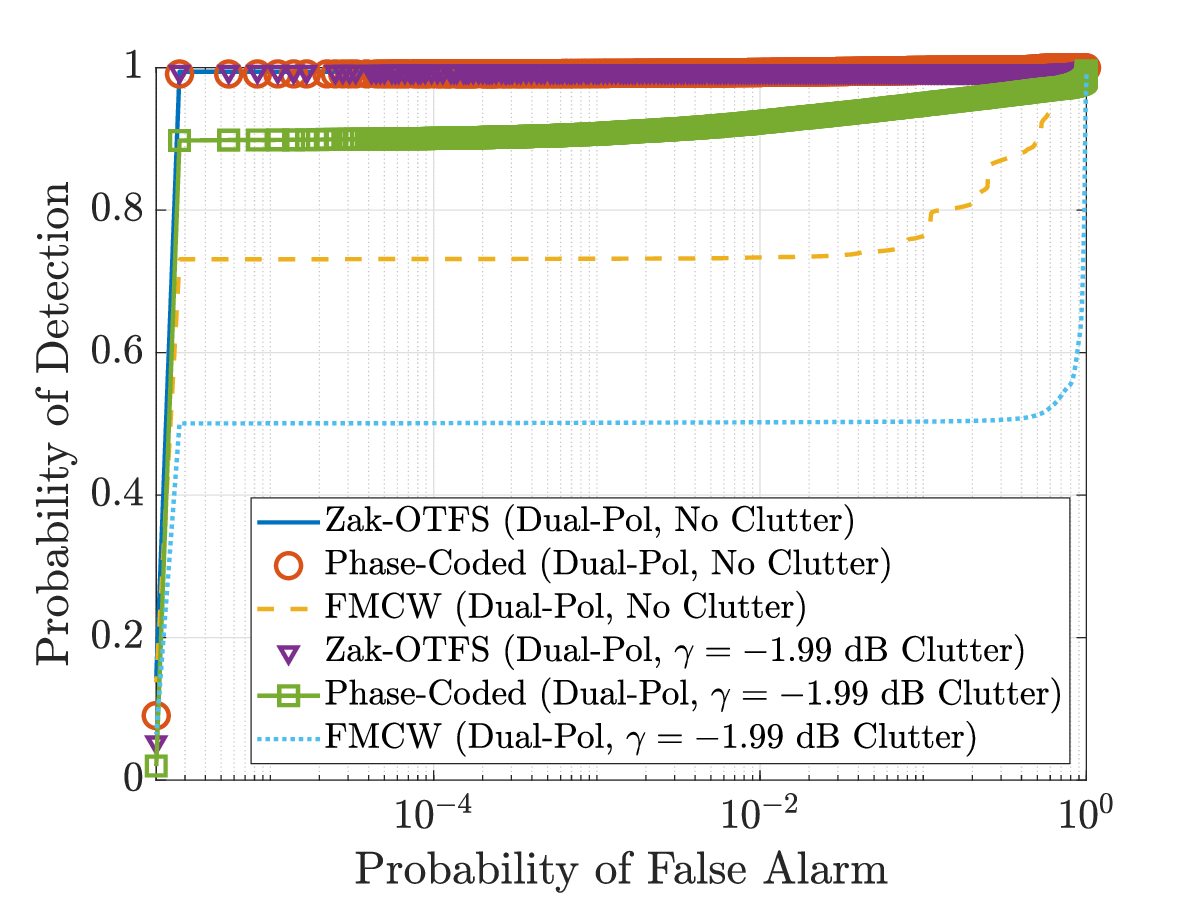}
    \caption{\textcolor{black}{Instantaneous polarimetry via Zak-OTFS exhibits greater resilience to clutter as compared to competing methods.}}
    \label{fig:clutter_roc}
\end{figure}

\textcolor{black}{Fig.~\ref{fig:clutter_roc} plots the detection ROC curve for single-target detection in constant-$\gamma$ clutter for all dual-polarized systems. For a metropolitan terrain and a carrier frequency of $f_c = 4$ GHz, we obtain the clutter parameter value $\gamma = -1.99$ dB from~\cite{Long1975_clutter}. Fig.~\ref{fig:clutter_roc} shows that instantaneous polarimetry via Zak-OTFS exhibits greater resilience to clutter as compared to methods based on phase-coded and FMCW transmissions.}

%% file: conclusion.tex
\section{Conclusion}
\label{sec:conclusion}

In this paper, we proposed an alternate approach for instantaneous polarimetry using the Zak-OTFS modulation. We designed a spread carrier waveform mutually unbiased to the Zak-OTFS carrier waveform, and proposed to simultaneously transmit both waveforms over orthogonal polarizations. Unlike existing methods with computational complexity quadratic in the time-bandwidth product, the proposed method enables instantaneous polarimetry at \textcolor{black}{near-linear} complexity \textcolor{black}{and greater clutter resilience}. \textcolor{black}{Future work will pursue experimental evaluation and applications to integrated sensing \& communication.}

% Via numerical simulations, we showed ideal polarimetric target detection and parameter estimation performance in the presence of noise. pursue algorithmic extensions that achieve the Cram\'er-Rao bound in the presence of clutter, explore the implications of the proposed approach in improving the reliability and spatial multiplexing capabilities of communication systems, and explore applications to integrated sensing and communication.